\documentclass[AMA,STIX1COL]{WileyNJD-v2}
\usepackage{lineno,hyperref}
\modulolinenumbers[5]
\usepackage{array}
\usepackage{mdwtab}
\usepackage{eqparbox}
\usepackage{url}
\usepackage{lscape}
\usepackage{rotating}
\usepackage{enumitem}   % For different enumeration numbering 
\usepackage{xfrac}
\usepackage{epsfig}
\usepackage{booktabs} % For Table formatting
\usepackage{tikz}
\usepackage{color}
\usepackage{multirow}
\usepackage{graphicx}
\usepackage{epsfig}
\usepackage{amsmath}
\usepackage{url}
\usepackage{float}
\usepackage{ragged2e}

\usepackage{algorithm}
\usepackage{algpseudocode}
\usepackage{pdflscape} % or {lscape}
\usepackage{multicol}
\usepackage{tabularx}
\usepackage{tabulary,longtable,afterpage}
\usepackage{supertabular}
\usepackage{balance}
\usepackage{wasysym}
\usepackage{pifont}
\newcommand{\cmark}{\ding{51}}%
\newcommand{\xmark}{\ding{55}}%

\graphicspath{ {figs/} }
\articletype{Research Article}%

\received{26 April 2016}
\revised{6 June 2016}
\accepted{6 June 2016}

\begin{document}

%\title{A Robust and Efficient Biometrics-Based Authentication and Key Agreement Scheme for E-Health Services}%\protect\thanks{This is an example for title footnote.}}
\title{Revenue Maximization Approaches in IaaS  Clouds: Research Challenges and Opportunities}%\protect\thanks{This is an example for title footnote.}}

\author[1]{Afzal Badshah}

\author[1]{Anwar Ghani*}

\author[1]{Ali Daud}

\author[2,3]{Anthony Theodore Chronopoulos*}

\author[4]{Ateeqa Jalal}

\authormark{Afzal \textsc{et al}}

\address[1]{\orgdiv{Department of Computer Science}, \orgname{International Islamic University}, \orgaddress{\state{Islamabad}, \country{Pakistan}}}

\address[2]{\orgdiv{Department of Computer Science}, \orgname{University of Texas, San Antonio}, \orgaddress{\state{ TX 78249}, \country{USA}}}

\address[3]{\orgdiv{(Visiting Faculty) Dept Computer Engineering \& Informatics}, \orgname{University of Patras}, \orgaddress{\state{26500 Rio}, \country{Greece}}}

\address[4]{\orgdiv{ Department of Computer Science}, \orgname{ University of Science \& Technology, Bannu}, \orgaddress{\state{Bannu}, \country{Pakistan}}}

\corres{*Dr. Anthony Theodore Chronopoulos and Dr. Anwar Ghani\\
\email{antony.tc@gmail.com, anwar.ghani@iiu.edu.pk}}

\presentaddress{Department of Computer Science, University of Texas, San Antonio, USA\\ Department of Computer Science \& Software Engineering, International Islamic University Islamabad, Pakistan}

% The paper headers
%\markboth{IEEE Survery ant Tutorial}

% =====================================================
%\maketitle

% === ABSTRACT =============================================
\abstract[Summary]{Revenue generation is the main concern of any business, particularly in the cloud, where there is no direct interaction between the provider and the consumer. Cloud computing is an emerging core for today's businesses, however,  Its complications (e.g, installation and migration) with traditional markets are the main challenges. It earns more but needs exemplary performance and marketing skills. In recent years, cloud computing has become a successful paradigm for providing desktop services. It is expected that more than \$ 331 billion will be invested by 2023, likewise, 51 billion devices are expected to be connected to the cloud. Infrastructure as a Service (IaaS) provides physical resources (e.g, computing, memory, storage and network) as VM instances. In this article, the main revenue factors are categorized as SLA and penalty management, resource scalability, customer satisfaction and management, resource utilization and provision, cost and price management, and advertising and auction. These parameters are investigated in detail and new dynamics for researchers in the field of the cloud are discovered.}


\jnlcitation{\cname{%
\author{A. Badshah}, 
\author{A. Ghani}, 
\author{A. Daud},
\author{A. Jalal},
\author{A.T Chronopoulos},
} 
(\cyear{2020}), 
\ctitle{Revenue Maximization Approaches in IaaS  Clouds: Research Challenges and Opportunities}, \cjournal{Transactions on Emerging Telecommunications Technologies, Wiley}, \cvol{2019;00:1--16}.}

\maketitle

\footnotetext{\textbf{Abbreviations:} SlA, Service Level Agreement; IaaS, Infrastructure as a Service;}

% ========= I. INTRODUCTION=====================================
\section{Introduction}
\begin{figure}[ht]
 \centering
	\includegraphics[scale = .7]{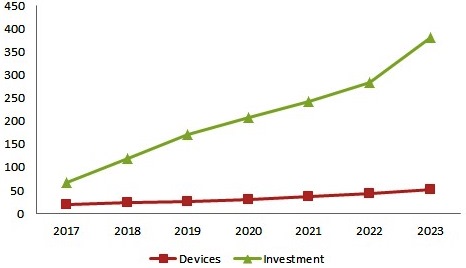}
	\caption{The cloud devices and revenue fore cost}
	\label{fig:forecost}
\end{figure}

The recent advances in smart technology generate a \emph{massive data} traffic. The 51 billion devices forecast  is a big number; even seven times greater than the whole world population \cite{Prop149}. These devices  will increase the annual size of the global data-sphere up to 175 ZB  \cite{Prop17,Prop154}.  Another report states, as shown in Figure \ref{fig:forecost} ,   that more than 331 billion dollars will be invested in cloud up to 2023 \cite{Prop17}. It needs special techniques and infrastructure to process the incoming data \cite{Prop04}. Furthermore, integrating Artificial Intelligence (AI) in smart devices makes the network more complicated. With this rapid development in smart technology, cloud computing is getting more and more attention. 

 Cloud computing  classifies desktop services into three primary categories:  \emph{Infrastructure as a Service (IaaS)},  \emph{Platform as a Service (PaaS)} and   \emph{Software as a Service (SaaS)} \cite{Prop07}. These services are provided in three different models:  \emph{private cloud, public cloud} and \emph{hybrid cloud} \cite{Prop08}.
 \emph{ Infrastructure as a Service (IaaS)} provides physical resources online (e.g., computing, storage, and networking). It provides servers, network connections, storage, and other related resources. Amazon Web Services (AWS) is most  popular  IaaS service provider  \cite{Prop19}.  Apart from AWS,  Micro Soft Azure, Google Cloud Platform, Ali Baba Cloud  and  IBM Cloud   are the leading IaaS providers in the market \cite{Prop123,Prop124,Prop125, Prop126}. 
\emph{Software as a Service (SaaS)} provides online applications (e.g., monitoring, finance, and communication)  to consumer, running on provider infrastructure. Oracle is one of the popular  SaaS service provider  \cite{Prop127}.  Apart from Oracle,  SAP , Cobweb , MuleSoftand SalesForce are the leading SaaS providers in the cloud market \cite{Prop128,Prop129,Prop130,Prop131}. 
\emph{Platform as a Service (PaaS)} provides online development tools (e.g, testing, analysis, and deployment services) for software development. Users and customers design software using programming languages, libraries,  and other tools.  Oracle Cloud Platform is one of the popular  PaaS service provider.  Apart from Oracle,  AWS, Google Cloud Platform, Microsoft Azure, and SalesForce  are the leading PaaS providers in the market \cite{Prop02,Prop03}.

\begin{center}
\begin{figure*}[ht]
	\centering
			\includegraphics[width=\textwidth]{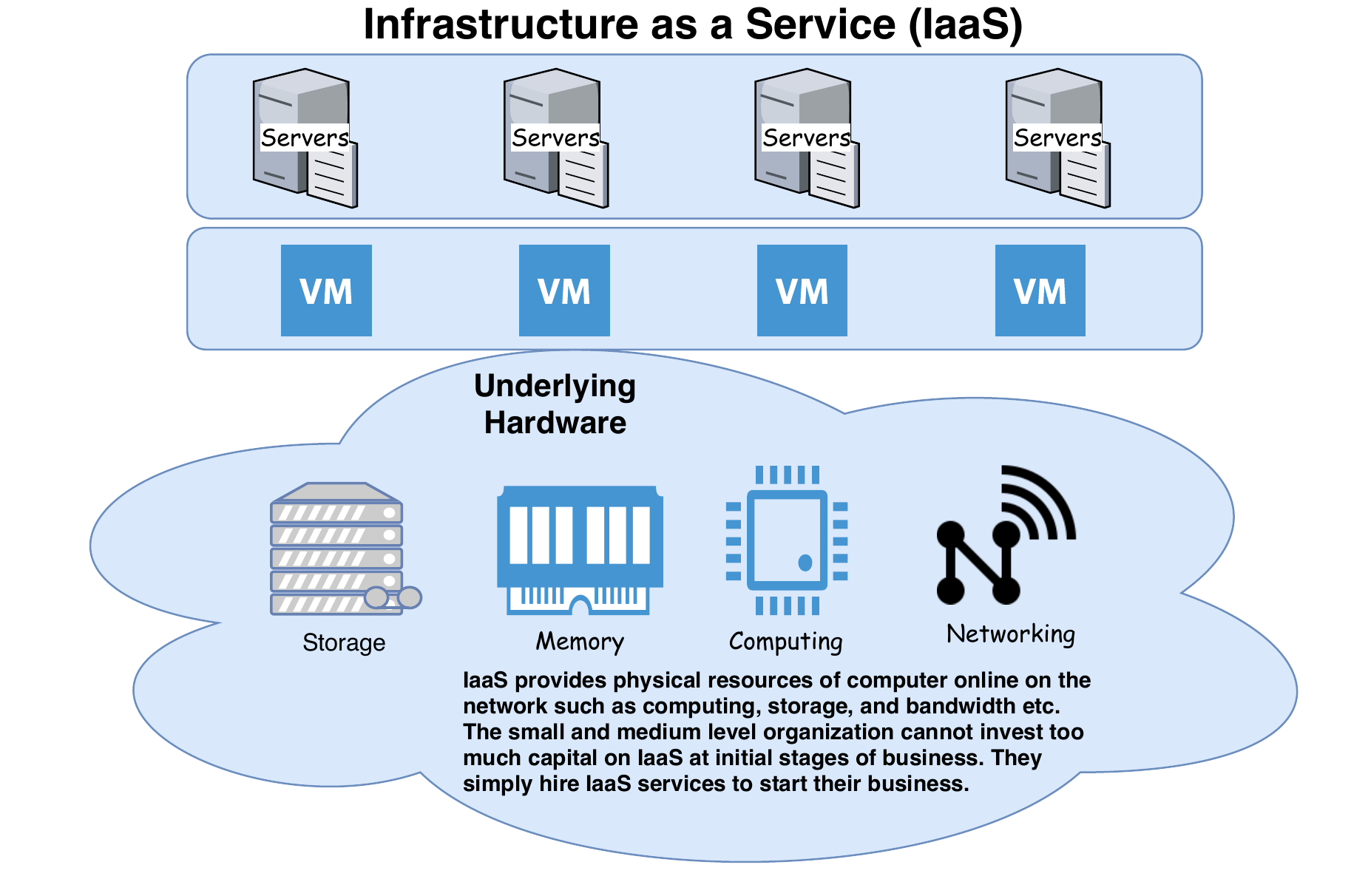}
		\caption{Infrastructure as a Service}
		\label{fig:iaas}
\end{figure*}
\end{center}

 \emph{Infrastructure as a Service (IaaS)} provides physical resources (e.g., computing, memory, storage, and networking) online as services. The traditional way of using physical resources has several limitations.   First of all, computer Infrastructure cost is high, secondly,  there are many issues in configuration, management, and maintenance \cite{Prop11}. Therefore,  small and medium level organizations cannot invest  capital on IaaS at the initial stages of business. They simply hire IaaS services to start their business.  IaaS clouds  are growing rapidly than other cloud services. The Compound Annual Growth Rate (CAGR) is 20.4 percent over the 2015-2020 forecast period.   Figure \ref{fig:iaas} shows the structure of IaaS clouds.
 
 \emph{IaaS utilization} is the primary method by which cloud business performance is measured and success is determined. In basic terms, it is a measure of the actual revenue earned by assets against the potential revenue they could have earned. In  IaaS clouds, virtualization, parallel and distributed processing techniques are used to improve the utilization. In virtualization,  single hardware is  shared  with many users. In parallel processing, many applications are run in  parallel simultaneously. In distributed processing, heavy workloads are processed on different servers \cite{Prop13}. Utilization plays a decisive role; in case of high utilization, the revenue increases otherwise resources remain underutilized and cannot be claimed in future and are wasted\cite{Prop14}.

\emph{Customer satisfaction} is a primary concern in business, which shows how much services are fulfilling the customers' needs. Customers are the measures of repurchases \cite{Prop145}.
Service \emph{performance} directly interferes the customers' satisfaction and providers' revenue. Cloud resources performance covers the number of parameters. It includes running time, waiting time, availability, reliability and security etc. These parameters thresholds are agreed during the SLA negotiation. On violation of these thresholds, defaulter pays penalties \cite{Prop99}.
\emph{Prices} play an active role in customers' satisfaction and attraction.  Where prices is directly proportional to performance, it is also inversely proportional to customers' satisfaction. Market needs such like pricing model which address the need of lower prices and higher performance customers \cite{Prop81}. 

Clouds is part of daily  life and have a big share in market, however,  there is still some resistance and complexities  towards providers revenue maximization. Due to these issues (e.g, customer dissatisfaction, cost, performance, penalties, under and overutilization etc), the providers' revenue is reduced. 

Relating to above issues,  \emph{extensive survey}  work  have been published and different parameters of cloud computing, such as performance  \cite{Prop116} , security  \cite{Prop118}, utilization \cite{Prop115}, fault tolerance  \cite{Prop119}, load balancing  \cite{Prop120} and resources provision  \cite{Prop121, Prop159},  have been discussed. However, revenue-maximizing factors have not yet been well investigated. Particularly, the classifications of the revenue maximization parameters are missing.  It needs further exploration for the academia and market.  

The  \emph{revenue maximization} depends on various parameters. In this survey, the key investigations, challenges and strengths  towards revenue maximization are reported. The associated factors are divided into seven different categories. All these factors contribute (directly or indirectly) to revenue maximization. 

This article  addresses research challenges arising from the following question:

\begin{quote}
\textit{With limited resources availability and heterogeneous customers’ demands, how to maximize the providers' revenue and performance by minimizing SLA violation and customers' dissatisfaction in IaaS clouds?}
\end{quote}

The main contributions of this article addressing this question are 
\begin{itemize}
\item  A chronological study of important studies of IaaS providers' revenue maximization by explaining their motivation, challenges and opportunities.
\item Describe the main influential features of revenue maximization in IaaS cloud. 
\item Classify the main features of revenue maximization into different categories 
\item Address the challenges and opportunities to bring to the attention of scholars in this area.
\end{itemize}

The rest of the paper is structured as follows. Section 2 reviews the  key factors to revenue maximization. Section 3 summarises the related work and categorizes it into seven different groups. Section 4 addresses the challenges and research direction and   finally, section 5 concludes the work.

\begin{center}
\begin{figure*}[ht]
	\centering
		\includegraphics[width=.8\textwidth]{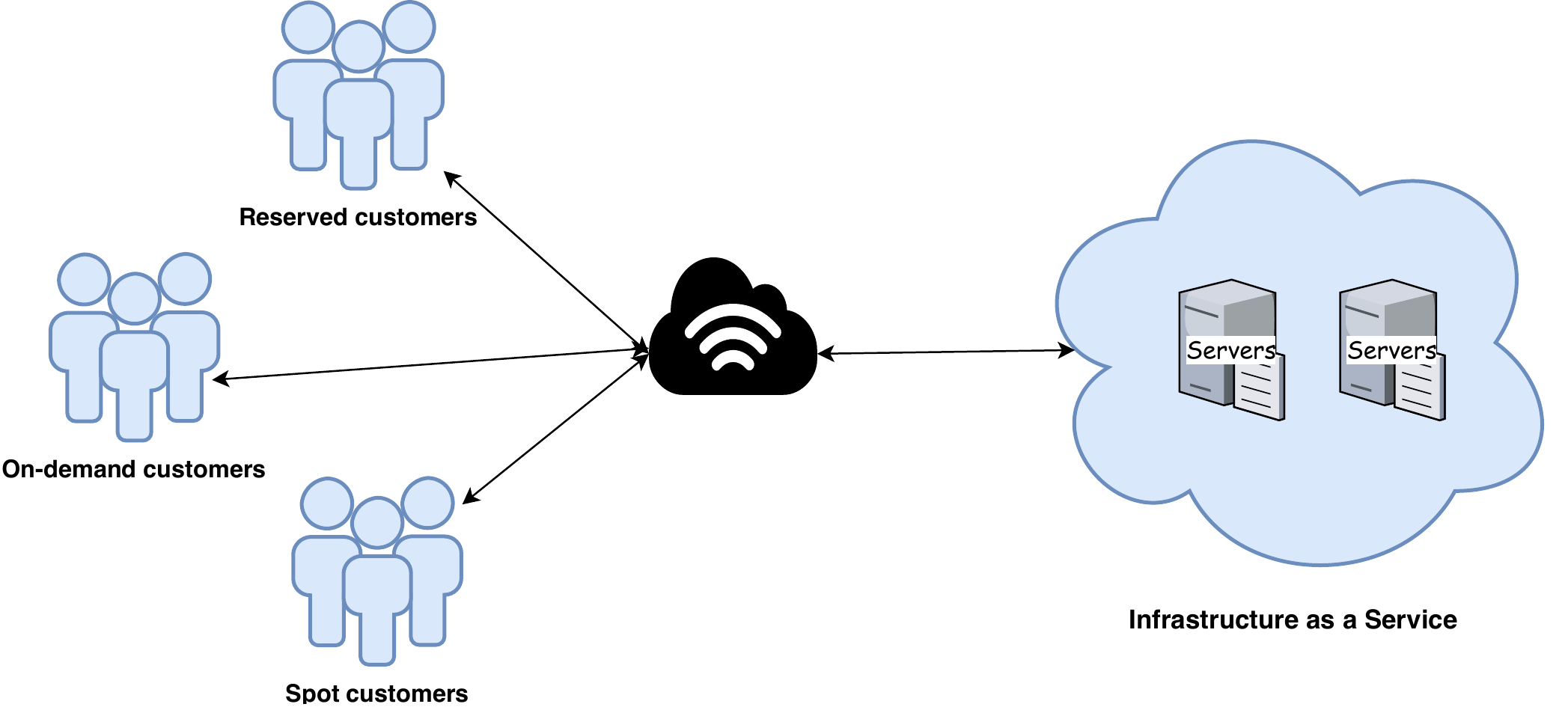}
		\caption{Customers of cloud computing}
		\label{fig:cust}
\end{figure*}
\end{center}

\section{IaaS revenue maximization}
Revenue is the prime focus of any business, particularly, in cloud computing, where there is no direct interaction between provider and consumer. Cloud computing provides services online,  engages more customers, reduces the cost and maximize the revenue.   Every provider wishes to generate good revenue and run the business up to the mark. In this study, the most effective parameters of cloud revenue are categorized as:  the performance of the services, SLA and penalties management, resources scalability,  resources utilization and scheduling, customers' satisfaction, cost and pricing management,  and advertisement and auction. This section discusses the preliminaries of these parameters in detail.

 \subsection{Performance management}
To keep the end-users satisfied, it is important to provide high-performance services. It is also challenging to convert the performance matrices to a quantitative form for measurement.  SLA implements the agreed performance parameters and imposes penalties in case of violations  \cite{Prop06}. 

\emph{Execution time}  shows the total time taken to execute the customers' workload.  This depends on the request type and resources which is to be executed. If the resources are not appropriate, it takes longer than usual.

\begin{center}
\begin{equation}
Per   \propto  1 / \tau_{run}   % Execution time
\end{equation}
\end{center}

\begin{center}
\begin{equation}
V_{n}    \propto  1/Per
\end{equation}
\end{center}

\begin{center}
\begin{equation}
\eta    \propto  V_{n} 
\end{equation}
\end{center}

The above expression shows that performance ($Per$) is inversely proportional to total running time ($\tau_{run}$). Further, the total number of SLA violations ($V_{n}$) is inversely proportional to performance. Wheres, penalties ($\eta $) are directly proportional to numbers of SLA violation  ($V_{n}$). These penalties have worse effects on cloud business. 

\emph{Response time} is the waiting time of customer request in the waiting queue. Response time depends on the underlying resources utilization. If the underlying resources are heavily utilized, it takes longer to execute new tasks. 

\begin{center}
\begin{equation}
\tau_{res} \propto  \upsilon \times 1/SS
\end{equation}
\end{center}
 
\begin{center}
\begin{equation}
Per   \propto  1 / \tau_{res}   % Execution time
\end{equation}
\end{center}

\begin{center}
\begin{equation}
V_{n}    \propto  1/Per
\end{equation}
\end{center}

The above expression shows that  response time ($\tau_{res} $) is directly proportional to total customer request ($\upsilon$) and inversely proportional to services scalability ($SS$). Further, performance is inversely propositional to  total response time ($\tau_{res} $).  Total number of SLA violation ($V_{n}$)   is  inversely proportional to performance, wheres, penalties  are directly proportional to number of SLA violation ($V_{n}$). 

\emph{Availability}  is defined as the presence of the agreed resources when they are required. Availability covers  these resources which are discussed in SLA.

\begin{center}
\begin{equation}
Avail \propto \frac{\tau_{avail}- \tau_{down}}{\tau_{com}}  % where com: commentment time
\end{equation}
\end{center}

\begin{center}
\begin{equation}
avail \propto 1/fail \times SS
\end{equation}
\end{center}

\begin{center}
\begin{equation}
\chi  \propto avail
\end{equation}
\end{center}

In the above expression $Avail$ shows the  availability of resources,  $\tau_{avail}$ shows the total availability, $\tau_{down}$ shows the down time,  $\tau_{com}$ shows the total agreed time. Furthermore, availability is directly proportional to resources scalability ($SS$) and inversely proportional to system failure ($fail$).  The cost  $\chi$ is  directly proportional to services availability.  

\emph{Resources reliability} is defined as the resources performing of the predefined functionalities for the agreed time under agreed terms and conditions. The resources are reliable if they are fault-tolerant and automatically recoverable.  Reliability also includes the fault tolerance, recover-ability and resources constancy. Lower reliability reduces customers' retention which leads to lower revenue.

 \begin{center}
\begin{equation}
Per  \propto Reliability
\end{equation}
\end{center}

\begin{center}
\begin{equation}
V_{n}    \propto  1/Per
\end{equation}
\end{center}

The above mathematical expression shows that reliable resources minimize the number of penalties. 

\subsection{SLAs and penalties  management}
Service Level Agreement (SLA) is an understanding, negotiated between a provider and a  consumer. Detailed Service Level Objectives (SLOs) are addressed, expected services, Quality of Service (QoS) and performance are agreed and approved  \cite{Prop04}. Both provider and consumer monitor services with agreed terms and conditions. If violations occur in agreed terms and conditions, penalties are imposed on provider  \cite{Prop08}. Clearly explained SLA improves the customers' satisfaction and guarantee the continuous provision of  services \cite{Prop09}.

SLA violation results in penalties that are applied in the form of lower prices during service failure or direct sanction.  Usually, cloud providers accept loaded SLAs, but later on, they cannot provide resources in accordance with the agreement. As a result, they have to pay a large portion of their income in fines. Performance, penalties, costs and revenues are complexly related. Their interdependence is explained in the following expressions.

\begin{center}
\begin{equation}
Per   \propto  Rev \times \chi \times  1/ \eta
\end{equation}
\end{center}

\begin{center}
\begin{equation}
Rev    \propto  1/ \chi
\end{equation}
\end{center}

The above comparison explains that an increase in performance ($per$) maximizes the revenue ($Rev$) and minimizes the penalties ($\eta$), but performance is also proportional to the cost ($\chi$), which is inversely proportional to the revenue. They have a complex correlation. The situation becomes more complicated if there is no proper framework, which clearly defines them. 

Recent advances in Information Technology (IT), attracted more transition towards cloud computing.  Furthermore, data ware and data mining techniques also attracts the market. Due to a large scale of data and internet business, it is  very essential to have clearly defined  SLA, otherwise provider will be disruptive with disastrous consequences in business. 

\subsection{Resources scalability}
Resource scalability is vital for QoS. Non scalable resources lead  to penalties and revenue degradation. Most of the performance parameters directly depend on the resources scalability \cite{Prop152,Prop158}.

\begin{center}
\begin{equation}
V_{n}  \propto  {\frac{1}{SS}} \times {\frac{1}{QoS}} \times {\frac{1}{Eff}}
\end{equation}
\end{center}

The above comparison explains the dependability of SLA violation ($V_{n}$) on services scalability ({SS}) , QoS, and services' efficiency (Eff). 

\subsection {Customers' satisfaction}
Customers' satisfaction is very important, which shows  fulfilling of customers' needs. Customers are the indicators of repurchase. It shows the point of differences. It is cheaper to retain existing customers than to bring in new ones. Usually, companies spend millions of dollars on customers' attention but a small investment on retaining them.

 \begin{center}
\begin{equation}
CS\propto SS \times Eff \times QoS
\end{equation}
\end{center}

Customers are the potential of any company.  A global survey by Accenture global customer's satisfaction report (2008) shows that prices are not the most important concern, the most important is the customers' satisfaction. Successful customer satisfaction services increase Customer Lifetime Value (CLV) which increases the company's revenue. According to McKinsey 13 \% unhappy customers tells about 9-15 people about their experiences. Customers’ satisfaction is most important for revenue and repurchases \cite{Prop21}. 

\subsection{Resources utilization and provision}
\emph{Services usage} is the most important method for evaluating the performance of assets and determining the success of the company. Basically, it is a measure of the real income generated by the assets in relation to the potential income they could have earned. The cloud uses virtualization, parallel processing and distributed processing to optimize the use of the underlying resources \cite{Prop140}. In the following equation,

\begin{center}
\begin{equation}
\displaystyle Rev \propto \rho \times \mu\times \nu
\end{equation}
\end{center}

\textbf{$Rev$ is }revenue earned, $\rho$ are prices, $\mu$ is resources utilization, $\nu$ is number of customers. 

Usually, in cloud computing, resources are reserved. If reserved resources are not used by reserved customers, they are  underutilized and wasted.These resources may be utilized with the permission of customers for higher revenue. This benefits both, provider as well as consumer. 
Resources utilization is expressed as

\begin{center}
\begin{equation}
\mu = \frac{Running (\sum_{i}^{n} VM)} {Available (\sum_{i}^{n} VM)}
\end{equation}
\end{center}

Different IaaS scheduling policies are used to allocate resources to different customers. Scheduling policies greatly affect the resources utilization, customers' satisfaction and providers revenue. 

\subsection{Cost and pricing management}
\emph{Pricing plans} play a very important role in generating revenue. The cloud market uses different types of pricing models. In the reservation pricing plan, customers reserve resources for a specific period, such as a month or a year. Resources are sold to customers with a reasonable discount. Customers pay the registration fee. In on-demand pricing, customers are billed individually. In this pricing system, prices are higher, however, providers are charged for breaching the SLA. In the spot pricing, prices are negotiated between customers and suppliers. Negotiation-based prices are used for underutilized resources. In differentiated pricing, cloud services are divided into different types of tier. Each tier has different prices. In unit pricing, customers are charged on a unit of space or bandwidth used. This pricing mechanism is more flexible than the tiered pricing mechanism. In the basic pricing of the subscription, customers are billed according to their subscription. Users receive a discount on early booking. The disadvantages of this pricing model are that the provider provides guaranteed services to customers and underutilization wastes resources. Usage-based pricing is also used by Amazon in which  customers are charged based on usage \cite{Prop21,Prop150}.

\subsection{Advertisement and auction}
Advertisement  spread positive prospective and  change negative impact. It attracts new customers and increases the utilization of under-laying resources. Auction is also used in cloud business to increase the resources utilization. Auction is usually used for the underutilized resources. Instead of wasting, auction gives some revenue.  It is very tricky because in case of SLA violation, penalties are imposed on provider party which minimize the provider revenue  \cite{Prop151}.

\begin{table*}[ht]
\begin{center}
\caption{Literature evaluation criteria}
\label{tab:criteria}

\newcolumntype{b}{X}
\newcolumntype{s}{>{\hsize=.1\hsize}X}
\newcolumntype{c}{>{\hsize=.4\hsize}X}
\setlength{\extrarowheight}{5pt}%
\begin{tabularx}{\textwidth}{s  c   b }

\toprule
\textbf{Symbol} &  \textbf{Criteria } & \textbf{Criteria Definition} \\

\midrule

C1  &   Performance management     &   In the fast and smart growing world, cloud providers are expected to provide outstanding performance. Customers buy the resources in expectation of good performance. Cloud performance includes the response time, running time, security, reliability and availability.   \\

C2  &  SLA  and penalties management     &      Not fulfilling the agreed QoS leads to SLA violation. Providers are charged for every SLA violation. SLA violation is not the cause of penalties only, it also creates customer dissatisfaction. A proper SLA violation framework saves the provider from penalties.      \\

C3  &    Resources Scalability    &     Cloud providers are supposed to handle customers around the world. Today smart technology is getting very fast and every person is going on smart devices. Providers handle this massive data if they have scalable resources. Otherwise, it leads to penalties and customer dissatisfaction \\

C4  & Customer Satisfaction       &  	Customer satisfaction is crucial for any business. Especially in the cloud, it is more important because there is no direct contact between providers and consumers. Customers' satisfaction means more workload for resources utilization.  Customers' satisfaction maximizes the revenue.     \\

C5  &    Resources' utilization and management   &   Proper resources scheduling and migration between Physical Machines (PMs) and Virtual Machines (VMs) keeps the performance up and saves the provider from SLAs' violations. Proper resources' utilization maximizes the revenue.      \\
	
C6  &    Cost and prices management   &     Cost and prices not only link to customers' satisfaction but it also critically affects revenue maximization. Cost is  minimized by a number of ways such as managing the employees, internal resources, security and electricity.  \\

C7  &    Advertisement and auctions  & 
Advertisement increases the number of customers, which increases the workload. Overutilization keeps the resources busy. Overutilization is tricky and risky. It may affect performance if not managed properly.    \\

\bottomrule
\end{tabularx}
\end{center}
%\end{small}
\end{table*}

\section{Literature  classification}
The literature on revenue maximization is diverse and does not depend only on a few parameters. Different authors have used different parameters and techniques for revenue maximization. In this survey, the literature on revenue maximization is classified (as shown in Table \ref{tab:criteria})  into performance management, SLA and penalties management, resources scalability, customers' satisfaction, resource utilization and provision, cost and prices management, advertisement and overutilization. For these parameters, research articles published between 2012 to 2019 are included in this survey.

To investigate the quality of existing literature,  a set of criteria is proposed in Table \ref{tab:criteria} to evaluate the work done so far on cloud computing revenue maximization.  Table  \ref{tab:Summary}   shows the detail summary of the related studies, Table  \ref{tab:analysis}   displays the comparative analysis,  \textbf{Table} \ref{tab:yearwise}   shows the year wise approach,  and Table  \ref{tab:pl}   discusses the limitation and potential of each study.

\subsection{Performance management}
Without good performance, services are useless in cloud computing market competition. No one can deny its importance. Investment in low performance services  will give no benefits. Such type of issues detract customers  rather than attracting them. The following research studies investigated the performance towards the revenue maximization.

Performance depends on different parameters. Sinung Suakanto and Saragih \cite{Prop22}  investigated the \emph{primary performance parameters}  in cloud computing.  They measured performance metrics using empirical methods.   Average response time and the number of requests time out of customers' requests were calculated in the cloud environment. Their results showed that an increase in the number of customers increases the average response time. Similarly, an increase in the number of users,   increases the request time out. This study considered the performance measurement of cloud computing and achieved  C1 criteria. 

Ran and Xi \cite{Prop59} advances the \emph{performance study}.   They worked on resource provisioning strategy with QoS constraints. The proposed framework used,  dynamic computing resources provision, cost-saving, and QoS guaranteed services.  An algorithm Service Provisioning Engine (SPE) and Request Queue Engine (RQE) were used to efficiently provide the resources. They tried to maintain QoS while minimizing the total cost. With performance, resources and cost management, this study achieved C1, C5 and C6 criteria (see Table I below). Danilo Ardagna and Wang  \cite{Prop97}  collected and analyzed the detailed literature about \emph{Quality of Service (QoS)}.The aim of their work was to study the QoS modeling area, categorizing contributions according to relevant areas and methods used. This study achieved performance management C1 criterion.

Over provision of VM degrades the performance. Underutilization also wastes resources. Kundu et al. \cite{Prop52} addressed this challenge by \emph{efficient resources allocation}. Resources were allocated dynamically. Three types of algorithms were proposed in this model. The first algorithm was MaxRevenue which searched  VM for maximum revenue. The second algorithm searched MaxGain and MaxLoss in all available VMs. The third algorithm compared the MaxGain and MaxLoss. They searched MaxRevenue and MaxLoss  VM. With these properties, it achieved performance management C1, resources scalability C3 and resources utilization and scheduling C5 criteria.

The same issue was further investigated by Feng and Buyya \cite{Prop50}. Revenue-oriented resources allocation was used for revenue maximization. Two types of solutions were proposed for revenue maximization:  (i) optimizing resources allocation and (ii) optimizing pricing mechanism. The main objective of their work was to find the proper allocation of servers among all instances to maximize the provider revenue. Two types of functions were discussed in this article: (i) Assurance Satisfaction Factor (ASF) and (ii) Response Satisfaction Factor (RSF). Both of these functions defined the achieved performance. In the pricing model, if agreed performance (ASF \& RSF) is achieved then customers are charged on regular price otherwise the provider is penalized and low prices are charged. With prices, performance and resources management, this study meets C1, C5 and C6 criteria.

Federation enhances the scalability, and results results in increased performance. Nazanin Pilevari and Sanaei \cite{Prop98}  explained the \emph{federated  Cloud Service Providers (CSPs)} to maximize the service quality and providers' revenue. They proposed an algorithm using an Integer Linear Program (ILP) to form the CSP federation.  They also proposed a heuristic-based algorithm for cloud federation formation following the ILP.  This study achieved performance management C1 and service scalability C3 criteria. 

Many studies contributed to improve this area. Koziris \cite{Prop108} proposed an approach to overcome transient cloud failures that happen during the application deployment. Apon \cite{Prop109} gave a systematic evaluation of Amazon Kinesis and Apache Kafka for the highly demanding applications. Bauer  \cite{Prop110} proposed a framework which keeps the performance up automatically. Gerndt  \cite{Prop111} proposed the auto-scaling performance evaluation for two-layered virtualization in cloud computing. Wang  \cite{Prop112}  proposed the Virtual Machine Placement Algorithm for high performance.

\emph{Delay balancing,} with today massive devices, is a hot issue. Gadey \cite{Prop115} investigated the energy consumption and delay balancing in IoT, Fog and cloud project.  They focused on two parameters, energy consumption and quality of service. They proposed an evolutionary algorithm to resolve this issue. 
The same issue was further studied by Duan \cite{Prop116}. They introduced a general framework for IoT, fog and cloud.  They proposed a delay-minimizing policy for IoT devices to minimize the service delay for IoT, fog and fog applications. 

These investigations and research  optimized the revenue and performance, however, there is a complex correlation between performance and revenue which is missing in the existing literature. For example, performance increases the customers' attention but on the other hand, this also increases the prices. Increase in prices, detracts the customers. Furthermore, with good performance, heavy workloads are expected.  This workload affects performance and SLA. All these queries need thorough examination and investigation.

\subsection{SLA and Penalties Management}
SLA is a \emph{contract} negotiated between a provider and a customer. Detailed service parameters are discussed and signed before starting the business. SLA creates a trusty relationship among business parties. If the signed parameters are violated, the defaulter is penalized. Penalties have worse effects on the provider side. With proper management, they may be minimized to maximize revenue.  

\emph{Efficient resources management} can save providers from SLA violation and penalties. Macas et al. \cite{prop31}  explored revenue maximization in cloud computing using Economically Enhanced Resource Manager (EERM). This has used bi-directional data between the market and tried to increase sales through dynamic pricing mechanisms. If the provider is unable to respond to the customer's request, a list of SLA violations is made. These service level agreements are violated, resulting in a smaller loss. Service level contracts that have lower revenues are also cancelled. If a VM is overloaded,  the workload is shifted to other VMs to reduce the performance reduction. This SLA model gives a good idea about increasing revenue but it also rejects high loaded SLAs, which creates un-trusty situations in business. With prices, resources and SLA management, this framework achieved SLA violation management C2, resources scheduling and management C5, and cost and prices management C5 criteria.

Wu et al. \cite{prop35} further studied the \emph{resource allocation} to avoid SLA violations. The focus of this study was the dynamic changing customers' demands with SLA. Services are provided according to SLA. The customers can also change their requirements. Providers try to increase the profit by ensuring QoS to broaden their business. The software is delivered in standard, professional and enterprise and accounts are created in a group, team, and department.  The contract is signed between a provider and a customer, if anyone violates, the defaulter pays penalties. The main objective of this work was to maximize the profit for the provider by minimizing the cost of VMs. By providing individual VMs to every user, no QoS degradation occurs, which minimizes the penalties. This paper achieved C1 for managing performance, achieved C2 and C3 by managing resources to avoid SLA violations. 

\emph{Automatic Service Level Agreement (ASLA)} was proposed by Christpher Redl and Schahram \cite{prop73}. It uses past knowledge, user requirements and job evaluation to automatically meet every SLA. Mapping is determined by the market participant. Before SLAs are made in the cloud market, customers and suppliers submit their SLAs. These templates include service level contract statistics, SLA parameters and service level objectives. In the market, users associate their private SLA with a public SLA that is closest to their needs. SLA assignment is used to assign two SLAs.  It automatically searches for similar SLAs on the market. ASLA lowers the costs of the market. With SLA and time management, this research resulted in C1 price and cost management and C2 SLA management. Automatic SLA saves time and money, but unfortunately, it is only an agreement. It does not guarantee the quality of the service and performance.
 
Investigations were limited to get all the cloud resources from any single point. Jennifer Ortiz and Balazinska \cite{Prop74} worked on a \emph{Personalized Service Level Agreement (PSLA)}. It acts as a broker between the service provider and  customer. PSLA rents out different types of services from different service providers and then offers them to other customers based on their needs. PSLA has solved this major issue. The user does not have to translate his requirements. They only upload their needs to the broker and he provides services according to the needs of the client. PSLA has solved the major problem of service provision, but it is a combination of different services and each service has a different service quality and service level. This enabled C2 criteria for service level management.

For SLA, it is important to define measurable parameters. Emeakaroha et al. \cite{prop40} advanced this idea by proposing  \emph{LoM2HiS (Low Metric to High-Level Services)}. It was part of the FoSII infrastructure. This SLA framework provides a platform for converting low-level statistics into high-level metrics statistics. This infrastructure contains the assigned metric repository and an agreed service level agreement. When a new client request arrives at the system, this infrastructure has assigned it to an assigned metric repository. LoM2HiS is an automatic framework for managing and maintaining service level agreements. This framework informs about future threats. With SLA and resource management, this study met the SLA and penalty C2,  and resources management  C3  criteria. This framework is the first step to measure the performance of cloud computing. However, it does not describe how these metrics should be measured or how they should be analyzed and integrated into a service level agreement for implementation.

 To analyze the SLA violations, Iyer \cite{Prop105} proposed analysis and diagnostic framework. Their study is based on 283 days of operational logs of the platform.  They received workloads from 43 customers, spread around  22 countries. They developed tools to analyze this workload. This study showed that about 93 \% SLA violation is caused by system failure.  This study achieved SLA and penalties management C3 criterion. 

Many authors and scholars contributed to  enhance the SLA in cloud. Shivani and Singh \cite{Prop100}  reviewed the detailed literature about \emph{SLA violation and its minimization}. Shahin Vakilinia and Elbiaze \cite{Prop101} proposed an integrated platform to detect and predict conditions where improving decisions are required.  They used neural networks to minimize SLA violations.  Their results showed that this improve web request response time by up to 7 \% and decreases SLA violation by 79 \% in the context of the web application. Gargouri \cite{Prop102} used advanced SLA management strategies to provide good quality services. This reasoning technique minimizes SLA violation. Alayat Hussain et al. \cite{Prop103}   proposed a Risk Management-based Framework for SLA violation abatement (RMF-SLA). This framework detects the SLA threat and recommended action is taken to avoid SLA violation.  Singh and Elgendy  \cite{Prop104} proposed three approaches, namely, gradient descent-based regression (Gdr), maximize correlation percentage (MCP), and bandwidth-aware selection policy (Bw), that can mostly reduce power losses and SLA violations. Jin  \cite{Prop106} proposed a privacy-based SLA violation detection approach for cloud computing using  Markov decision process theory.Zhonghai \cite{Prop107}  worked on availability commitment in cloud computing to minimize SLA violations. 

\emph{SLA} and \emph{penalties} are deeply investigated.  Different mechanisms are reviewed to minimize the SLA violations. Penalties are also well investigated to keep this burden minimum. The main drawback which needs improvement that most of the provider cancelled the SLA as their workload increases. Most of the provider with limited resources also have admission control and heavy loaded workloads are cancelled. These issues need further exploration.

\subsection{Resources scalability}
\emph{Resources scalability} is directly proportional to performance, which is directly proportional to revenue. With scalable resources, more customers are entertained with excellent performance. With non-scalable resources, performance decreases which leads to SLA violations. Therefore, for the cloud business, resources must be scalable. 
 
\emph{Scalable resources} are directly proportional to revenue generation.  For resources scalability, Gao et al. \cite{Prop38}  proposed Cloud Bank Service Level Agreement (CBSLA). In CBSLA, services are used by the service provider and these services are stored in a service pool. Two types of SLAs are used. The first SLA is signed by the service provider and the cloud bank, while the second is signed between the service consumer and the cloud bank. Cloud banking works as a cloud service broker. Various SLAs are being negotiated with various cloud service providers and customers. Pooling services is also a difficult task, so their use and implementations are very complex. This investigation achieved the SLA management C2  and partially resources management C5 criteria.

\emph{Insourcing} and \emph{outsourcing} is the first step to deal with resources scalability. Hadji and Zeghlache \cite{Prop43}  utilized these techniques in federations. The provider uses outside federation only when its cost is lower than internal cost, and also insourcing is used when internal utilization is lower. Mathematical programming approach is used to do good outsourcing and insourcing decisions. For minimum cost and maximum revenue, four possible actions are discussed.  An optimal number of machines are activated for any request. On maximum utilization, some requests are outsourced. In limited utilization, some of the internal resources are insourced.  Nodes which are not in use, are turned off to save power. The main limitation of this framework is that insourcing and outsourcing can be done only with registered providers. With insourcing and outsourcing capabilities, this study achieved resources scalability C3 criterion.  

\emph{Resources migration} is one solution to handle the overutilization issue. Upadhyay and Lakkadwala \cite{Prop48} advanced the resources migration in cloud computing. Migration is used in distributed systems, when data and applications are transferred from overloaded systems to underutilized systems. Usually, two types of migrations are used, primitive migration and non-primitive migration. The proposed framework used two types of algorithms. The first algorithm checks the overloaded and underutilized systems. Only if the target system has enough space to run the burden of the overloaded system, the workloads are transferred to it.  The second algorithm works for effective resources allocation in the cloud system. With migration property, this study achieved resources scalability and management C3 criterion. 

Migration work was further advanced by Li et al. \cite{Prop44}. A framework was proposed to migrate data to other systems. Migration may be as a whole application work, partial migration, component replacement or codify. Those methods need different architecture and environment to migrate the data. They used the Eucalyptus platform to evaluate their framework. This study discussed the cost and prices C6 and resources scalability C3 criteria.

Mobile cloud is getting a small share in utility computing. Their reliability increased with the usage of fog computing.  Samanta and Chang \cite{prop83} investigated mobile cloud revenue. It is not possible to perform every application task on mobile because it requires huge  space and memory. They have designed an approach to make some of the cloud sources and some sources work on the peripherals. They proposed an adaptive release system for Mobile Edge Computing (MEC) services to maximize total revenue. The execution of certain processes on the edge and some on the cloud server influences the performance due to network delays. They consider the delay-sensitive and delay-tolerant edge services by designing an offloading algorithm. By migration task between edge devices and a cloud server, this study achieved resources scalability C3 criteria. A challenge with this scheme is that migration of live data between edge devices and cloud servers decreases the performance. It may also create security issues.

User distance from server decreases the performance and increases the delay time. Hou Deng \cite{prop84}  addressed this gap to maximize the revenue of \emph{geographically distributed} data centres.  The solution for this is to build geographical data centres but It needs too much investment to build new data centres. The authors proposed a solution for this issue to hire geographical resources from universities or other institutions to process the data locally. This minimizes the cost of the provider. The challenge with this approach is that getting geographical resources may need many SLAs as per the region. It is also not possible to hire these resources around the world. Secondly, it may have worse effects on performance. With cost minimization, this study achieved C6 criterion. 

As per the above discussion, authors offered different solutions to handle the \emph{resources scalability} challenges. Federated cloud partially overcome this issue by sharing resources within the union. However, the drawback of this proposal is that providers are compelled to hire resources from particular providers with fix rates. Another study suggests that workloads should be accepted from only the surrounding area. This study has some positive directions that with a lower distance, performance of the system may increase, but this shrinks the concept of cloud computing. All these questions need to be addressed and require further focus  in  future research.

\subsection{Customers' satisfaction}
In any business, \emph{customer satisfaction} is the top priority. How good quality and scalable resources do providers have but if customers are not satisfied, revenue may not be earned. Customers' satisfaction is important in any business but in cloud computing, it is much important because there is no direct communication between providers and customers. Customers' satisfaction is based on the performance parameters discussed in the performance management portion. If these parameters are achieved with agreed SLAs,  customers will be satisfied. The following research studies discussed customers' satisfaction and classification in the cloud computing business. 

Customers' satisfaction is hard to deal with in terms of measurement and keeping them satisfied. Nazanin Pilevari and Sanaei \cite{Prop98}   developed conceptual criteria to \emph{measure customers' satisfaction}. These criteria were based on previous studies and expert opinions. The criteria consist of security, efficiency and performance, adaptability and cost.   Secondly, they developed a fuzzy logic, both, human and machine to fulfill the above criteria. With the customer satisfaction study, this study achieved  C4 criterion. 
R. A. Asaka and Ganga \cite{Prop99} extended the customers' satisfaction to Software as a Service (SaaS) cloud.   This model followed the survey and statistical analysis of client accounts from one of the world´s largest IT companies.  According to their findings, quality of the execution, quality of the implementation, and relationship are factors with higher influence on client satisfaction. They achieved customers satisfaction management C3 criterion. 

With limited resources, it is hard to satisfy customers. Dividing customer satisfaction parameters into different layers may ease the work of the providers. Hamsanandhini and Mohana \cite{prop65}  \emph{categorized} all clients into different groups.  They used different policies to maximize revenue. The policies selectively violate the SLA.   Overselling resources,  hybrid pricing,  booking already used resources, and client priority is artificially added. Clients are classified on different parameters such as the clients' relation to the provider and quality of service required by the customers. With the client classification framework, this study achieved customer satisfaction C4  criterion.

\emph{Customer classification} was further studied by Huu and Tham \cite{Prop95}. They worked on SLA enforcement by client classification. They introduced a set of policies to manage SLA during its operations. They classified the clients according to their affinity and QoS. According to these policies and classification, high-priority clients are selected.  According to their classification services are provided. They achieved performance management C1 and customer satisfaction C4 criteria. 
Manzoor et al. \cite{prop64} worked on \emph{customers centred approach} for IaaS cloud. The proposed framework works in three phases. In the first phase, customers submit their requirement specification to the Cloud Service Providers (CSP) and providers start service provision. In the second phase, services are monitored according to Cloud Information System (CIS). In the third phase, monitoring reports are compared with CIS. With a customer-centred approach, this framework achieved C4 criterion. 

Mei et al. \cite{prop63}  discussed two customer satisfaction parameters, which affect the revenue most, \emph{Quality of Services (QoS)} and \emph{Price of Services (PoS)}. QoS shows the expected performance and PoS shows the comparison between the predefined price and the actual price. They developed a model which optimizes the QoS and prices for customers’ satisfaction. With customers’ satisfaction, the number of customers increases which increases the revenue of the provider. With QoS,  customer satisfaction, and prices management this work achieved C1, C4 and C6  criteria.   

The above studies investigated customer satisfaction challenges and suggested different frameworks to satisfy customers. The main contributions of these studies are to classify the customers into different layers and according to these layers, providers create a customer satisfaction layer. Questions about what to do in case of lower resources with higher workloads and also to optimize the performance and prices still remain open. These need further investigations to optimize performance and prices and also to handle massive resources with limited resources.

\subsection{Resources utilization and provision}
Cloud resources are not storable. They are wasted, if not utilized  in time. Resources utilization extend the cloud provider business. For proper resources utilization, it is necessary to have a good utilization and scheduling framework. The following studies investigated these cores to increase provider revenue.

Cloud resources need to be utilized on time. Similar to other utilities (e.g. power or water), cloud resources cannot be stored to be used later on. To fill this gap, Shin et al. \cite{prop51}   proposed an algorithm which enhances \emph{deadline guaranteed resources utilization}. All jobs are sorted according to their arrival time, each job worse execution time is calculated. All VMs resources information is sorted in a Cloud Information System (CIS). VMs are allotted to different jobs using worse case execution time and deadline sorting. With these properties,  this research achieved resources scheduling and management C5 criterion.  

In cloud computing, resources and workloads are geographically distributed. In this situation, it is very difficult to perfectly match virtual machines with different workloads. Balagoni and Rao \cite{prop62} worked on the task planning policy in a heterogeneous cloud environment. This study investigated the locality predictor that increases the matching factor and the performance of cloud computing. They developed an algorithm that worked as basic functions on location, and load  prediction.With these characteristics, this study achieved resources management and scheduling C5 criteria. 

The previous studies stressed admission control, however,  Yuan et al \cite{prop66}  proposed \emph{Profit Maximization Algorithm (PMA)} with delay tolerance. They proposed temporal task scheduling for profit maximization in hybrid clouds. They addressed the problem, handling all the incoming tasks with limited private cloud computing resources. Private cloud workloads are scheduled to the hybrid cloud. The temporal task scheduling algorithm allows running the private task on the private and public cloud. With scheduling properties, this work achieved the resources scheduling and management C5 criterion. 

Ibrahim et al. \cite{prop71}  worked on \emph{task scheduling} in cloud computing. They proposed an enhancing task scheduling algorithm, which calculates all available resources and task request for processing. Groups of users are allotted to different VMs according to the ratio of needed power. With resources scheduling algorithm, this study achieved the resources scheduling and management C5 criterion.

\emph{Live cloud migration} was utilized by Mansour et al. \cite{Prop45} .  This work is divided into three phases. In the first phase, permission is granted to every VMs for migration. In the second phase, the required information for resources is gathered to decide either to migrate the resources or not. In the third phase resources utilization are monitored to avoid overutilization. This study discussed the resources scheduling C5, and cost and prices C6  criteria. 
Santikarama and Arman \cite{Prop46} developed an architecture framework for non-cloud to cloud migration. They used Economic Customer Relationship Management (ECRM) to efficiently migrate the data.  This study achieved the resources scheduling  C5 criterion.

Live migration was further investigated by Tsakalozos et al. \cite{Prop47}.  They proposed a framework which reduces SLA violation by migrating the resources on time. They proposed a scalable and distributed network for customers. The migration is done within time windows. This study achieved resources scheduling  C5 and SLA management C4 criteria.   
Gao et al. \cite{Prop38}  worked on \emph{transcoding} in the cloud for profit maximization. Transcoding is widely used for online video streaming. They proposed time scale stochastic optimization framework to maximize provider profit and also service performance. Transcoding in normal condition, waste about 30 \% resources and time. Cloud-based transcoding is a new way, which saves time and resources. In cloud computing, with numbers of VM availability, parallel transcoding is used, which greatly reduced resources and time wastage. This work achieved resources scheduling C5 criteria. 

The above review explains the optimum utilization of cloud resources, its challenges and suggested different frameworks to maximize resources utilization. The main challenges toward resources utilization are admission control and SLA violation. Providers do not overload their resources due to the fear of  SLA violation.  These complexities need further investigations to optimize resources utilization and SLA violation. 

\subsection{Cost and prices management}
Prices have direct effects on customers. Some customers do not care about prices but want high performance, some customers do not care much about performance, they are not able to pay high prices. Therefore, there must be a framework, which will manage the prices according to customers’ requirements. 

For better utilization and to increase the providers' revenue, nowadays a dynamic pricing mechanism is used for dynamic customers' requests. Ran and Xi \cite{Prop59} investigated a model to increase the revenue of the cloud computing provider.  E.g. Amazon EC2 is also offering dynamic pricing since 2009. Dynamic pricing mechanism in IaaS causes  many problems. They formulated a program which deals with such problems and handles infinite horizon cases. This study worked on resources scalability C3, and pricing C6 criteria.

\emph{Market analysis} plays a good role to prepare the resources according to the incoming demands. Zhang and Boutaba \cite{prop49} investigated a model to maximize the revenue of cloud providers. The market analyzer is used to analyze the market incoming request briefly. Then, they are using capacity planner which prepare the machines and resources capacity according to the reports of a market analyzer. This model is using both price mechanisms, dynamic and static. Different algorithms are used to predict the situation to use suitable prices mechanism technique which would be suitable for certain situations. This study discussed the cost and prices management C6 criterion.

\emph{Admission control}  is  also discussed in cloud literature to maximize the providers' revenue. Toosi et al. \cite{prop42}  worked on optimizing admission decisions to accept only those contracts whose revenue is higher. In the proposed model three types of pricing mechanisms are used to maximize provider revenue in limited resources availability. These pricing plans are spot market, on-demand,  pay as you go and reservation.  Two types of algorithms are used in this model.  Reservation contract is applied first and then the remaining capacity is utilized using spot instances. Revenue is earned from the upfront reservation, revenue from reserved, on-demand and spot instances respectively. Live reservation and running on-demand SLAs are kept within the provider capacity so that to control SLAs violations. With customer classification and capacity planner,  this study achieved  C4 and C5 criteria.

To \emph{optimize the resources utilization with prices}, Chi et al.  \cite{prop67}  proposed profit maximization using pricing methodology in cloud infrastructure. They worked on efficient resources utilization and pricing models to increase the number of customers. As customers' requests are accepted,  they can easily and fairly be fulfilled. Higher pricing is used for those customers, whose jobs are difficult to fulfil. Two steps are used for pricing calculation, unit price redistribution and revenue redistribution. This study achieved resources utilization and management C5 and cost and prices C6 criteria. 

\emph{Cloud cost} was discussed by Zhou et al. \cite{Prop11}. They worked on cost optimization in IaaS clouds.  Dyna, a scheduling system was developed,  to minimize the monetary cost. A (*) based search transition is used in this framework to search best price VMs. Finding the best price VMs maximizes revenue. This study achieved the cost and prices C6 criterion.    
Xu and Li \cite{Prop70} worked on \emph{hybrid cost} and priority-based scheduling in the cloud environment. In this framework, they proposed a new hybrid economic algorithm which takes both the cost and priority scheduling to maximize the resources utilization. With scheduling and prices management they achieved C2 and C6 criteria.

As profit is directly affected by costs, Zhao et al. \cite{prop56}  tried to minimize the cost of the resources to maximize the provider profit.   Furthermore, they worked on individually fulfilling the objectives. Their objective A is to minimize the cost, objective B is to start the queries execution at the earliest time and objective C is to combine objective A and B. With cost, SLA and resources management characteristics, this study achieved C5 and C6 criteria.

About 80 percent of all the power of data centers are consumed by the server. Power consumption is the major consumer of cloud revenue. Tevi Yombame Lawson \cite{prop80}  proposed an economic framework for resources management. This proposed framework minimizes the usage of power.  They proposed an On-Off model for servers to save power and to maximize profit. The total power consumed by the Data Centers is $\propto$  $\ast$ p, where $\propto$ is the total CPU cores and P is the power consumed during the extreme time. Power consumption reduction, minimizes the costs. With cost minimization,  this study achieved C6 criterion.  

Tang and Chen \cite{prop69} worked on \emph{pricing and capacity planning}. They discussed two types of models, monopoly IaaS providers market, and multiple IaaS provider market. The optimal solution is searched in dynamic and static pricing for profit maximization. With prices and capacity planning properties, this study achieved the resources scheduling and management C5 and cost and prices management C6 criteria.

Mehiar Dabbagh and Rayes \cite{Prop81} tried to answer two main questions,  i) where to place the incoming workload? and ii) how many resources should be allocated to this workload? This decision matters much in profit maximization because wrong placement wastes the resources and delays the tasks. A challenge with this framework is that running too many workloads only on a single machine may decrease the performance which leads to penalties.  With resources and cost management, this study achieved C5 and C6 criteria. 

Sharing common resources among VMs reduces the cost.  To advance this idea, Rampersaud and State \cite{prop82} worked on Sharing Aware Virtual Machine Revenue Maximization (SAVMRM) problem.  They used a greedy algorithm to share the common memory and resources among the VMs hosted on one physical machine.  This algorithm result shows a great deal toward revenue maximization.   Sharing resources among different VMs may cause security and performance issues. With cost management discussion, this study achieved C6 criterion.

Authors in \cite{Prop93}, worked on the novel revenue optimization model to address the operation and maintenance cost of the cloud servers. Authors used an algorithmic and analytic approach to solve the issues of optimal utilization of the resources. These algorithms  minimize the power and operational cost to maximize the profit. With cost management discussion, this study achieved C6 criterion.

Lower prices attract customers, however, it also creates performance issues. The above studies  investigated  how to optimize the cost, prices and performance, however, this needs further investigation to optimally determine these parameters. 

\subsection{Advertisements and auctions} 
\emph{Advertisement} attracts more customers and obviously, it means more utilization. Overutilization is another technique to increase the underutilized resources' utilization, however, an issue is that it may lead to SLA violation. This may decrease performance and also customer satisfaction. 

\emph{Over-commitment} of resources is a complex decision. This increases the resources utilization but in case of risk miss calculation, it  also increases the SLA violation. Dabbagh et al.  \cite{prop57} worked on resources utilization through cloud resources over-commitment. They used over commitment for minimizing  Physical Machine (PM) overload and minimizing the number of VMs. In over-commitment, instead of initializing new VMs and to migrate the overloaded resource, simply resources are transferred to the PM. This saves VMs migration and initialization costs. The proposed framework uses different types of predictor such as VM utilization predictor and over- loaded predictor to increase resources utilization. With cost management and resources overutilization, this study achieved C6 and C7 criteria.  

Metwally and Ahmed \cite{prop58}  worked on  IaaS \emph{resources allocation}. The main problem with cloud service providers is to handle a large number of requests of IaaS customers. Authors proposed Integer Linear programming technique and a mathematical programming model to find the optimal solution. They used large scale optimization tools and column generation formulation to allocate resources in a large data-centre. They worked on resources over-utilization C7 criterion.

\emph{ Auction} can attract customers. Samimi and Mukhtar  \cite{Prop148} proposed combinatorial double auction model for revenue maximization. They extended already two models for double auction. The proposed model uses different phases and algorithms to maximize provider revenue. In the first phase, the cloud provider advertises its resources to the Cloud Information System (CIS). Every broker gets information from the CIS. The second phase generates the resources bundles according to user requirement; thereafter, auctioneers are informed. In the fourth phase winners and losers  are determined. In  the fifth phase resources are allocated to the winners. In the sixth phase, the payable amount is determined by the price model. With these properties, this study achieved resources auctioned C7 criterion.  
Hammoudi et al. \cite{prop60} worked on \emph{load balancing} in cloud computing.  With load balancing and parallel processing, large tasks are completed within a short time limit. They implemented this platform in the JADE platform. With load balancing characteristics, this study achieved C7 criterion. 

The above investigation shows that advertisement increases the numbers of customers. People also take interest in the auction and it attracts more customers. More customer may overutilize the resources which may lead to SLA violation. These complexities need further research and exploration.

\begin{longtable}{p{3cm}p{9cm}p{5cm}}
\caption{Detail summary of related studies}
\label{tab:Summary} \\

\toprule
\textbf{Paper and Authors } & \textbf{Major Contribution}  &  \textbf{Limitations}  \\ 
\midrule

\endfirsthead
% intestazione normale

\toprule
\textbf{Paper and Authors } & \textbf{Major Contribution}  &  \textbf{Limitations}  \\ 

\midrule
\endhead

\midrule

\endfoot
% piede finale
\bottomrule

\endlastfoot

Kundu et al. \cite{Prop52} & Efficient resources allocation to dynamic requests for revenue maximization. MaxGain,  MaxLoss and MaxRevenue algorithms were used to select the best economically  VMs. Incoming requests were run on globally selected VMs  &  Over provision of VM  violates SLA and decreases services performance. Performance badly affects incoming customers.     \\

Macas et al. \cite{prop31} &  Economically Enhance Resource Manager (EERM) is used for revenue maximization. In extreme utilization,  those SLAs are rejected whose penalties are lower. & Rejecting existing customers and not accepting heavy loaded SLAs have very bad long term impact on business.  \\ 

Gao et al. \cite{Prop38} &  Cloud-based transcoding is a new way, which saves time and resources. In cloud computing, with numbers of VMs availability, parallel transcoding is used, which greatly reduced resources and time wastage     & This framework is only for  live video  transmission and cannot be utilized in other data.  \\

Hadji and Zeghlache \cite{Prop43} & Insourcing and outsourcing were investigated in the federation to maximize the cloud provider's revenue. The provider used outside federation only when its cost was lower than internal cost, and also insourcing was used when internal utilization was lower & They did not discuss efficient algorithms for insourcing and outsourcing. Insourcing and outsourcing minimize performance. \\ 

Amit et al.  \cite{prop83} & An approach was designed to run some of the resources on the cloud and some resources on the mobile devices. They proposed an adaptive service offloading scheme for Mobile Edge Computing (MEC) to maximize the total revenue.  &  A challenge with this scheme is migration live data between edge devices and cloud server may decrease the performance. It may also create security issues.     \\

Hou Deng \cite{prop84} & A solution was proposed for  distance users issue to hire geographical resources from university or other institutions to process the local data. This minimizes the cost of the provider by decreasing the delay time.  &  The challenges with this approach are: getting geographically resources and the SLAs managing as per each region.    \\

 Hamsanandhini and
Mohana \cite{prop65} &     Previous work focuses on admission control but they proposed Profit Maximization Algorithm (PMA) with delay tolerance. The workload was scheduled from private cloud to hybrid cloud.  &  How to handle these customers with limited resources is still a big issue.    \\

Mei et al. \cite{prop63} &  Customers' satisfaction was explored for revenue maximization. A model was developed which optimized QoS and prices for customers’ satisfaction. With customers' satisfaction, the number of customers increases which increase the revenue of the provider.  & Customers' satisfaction increases the customers, but how to handle these customers' workload with limited resources, is still an issue.  \\ 

Toosi et al. \cite{prop42} &  Optimal capacity was utilized to maximize the providers' revenue.  Only those SLAs are accepted, whose revenue is higher. For customers' attraction, they used different pricing schemes. &  Those customers are rejected, whose revenue is lower. Rejection of customers has a very bad impact on business. \\ 

Zhao et al. \cite{prop56} &  This study investigated individually fulfilling of the objectives. Their objective A was to minimize the cost, objective B was to start the queries execution at the earliest time and objective C was to combine objective A and B. They proposed profit optimization algorithm. & They did not discuss how to entertain heavy loaded SLAs, and what to do in extreme utilization.  \\

Tevi Yombame Lawson \cite{prop80}  &  They proposed an On-Off model for servers to save power for revenue maximization. Only limited PMs are turned on to meet the customers' requirements. Power consumption minimizes the costs & Switching off some servers increases the workload on other servers. This may affect the performance of the services. \\

Mehiar Dabbagh and Rayes \cite{Prop81} &  They worked on cloud profit's maximization by efficient resources allocation, costing and pricing.  This maximizes the utilization of a single physical machine instead of running many physical machines for the work having a capacity of a single physical machine.  & A challenge with this framework is that running too many workloads only on a single machine may decrease the performance which leads to penalties.      \\

Rampersaud and State \cite{prop82} & They used a greedy algorithm to share the common memory and resources among the VMs hosted on one physical machine. Sharing common resources among VMs reduces the cost.   &  Sharing resources among different VMs may cause security and performance issues.   \\

Snehanshu Saha and
Roy \cite{Prop93} & They worked on a novel revenue optimization model to address the operation and maintenance cost of the cloud servers. They used an algorithmic and analytical approach to solve the issues of optimal utilization of the resources. These algorithms and analysis minimize the power and operational cost to maximize the profit.   & The challenge with this  framework is that performance increases the service cost. \\

Hong Xu \cite{Prop78} & Dynamic prices were utilized to attract more customers. Resources utilization increases providers' revenue.  &  More customers require salable resources. No solution was discussed for penalties and heavy loaded SLAs. \\ 

Adil Maarouf and
Haqiq \cite{Prop75} & SLA's penalties functions, strengths and weakness were explored.  A novel penalty framework was proposed for calculating the penalty of the violations and presented a formulation for this penalty definition. &  Only a penalty framework is discussed. Further exploration is required to minimize the penalties. \\	

Qi Zhang \cite{Prop77} & They developed a framework which uses a market analyzer and capacity planner to maximize the providers' revenue. Capacity planner prepares the machines and resources capacity according to the reports of a market analyzer. & Rejecting existing customers and not accepting heavy loaded SLAs may have very bad long term impact on business.     \\ 

Wu and Buyya \cite{Prop34} &  This study focused on the dynamic changing customers' demands and QoS concerning SLA. The services are delivered in standard, professional and enterprise and accounts were created in a group, team, and department. This attracted more customers. & Rejecting existing customers and not accepting heavy loaded SLAs have very bad long term impact on business.     \\ 

Afzal Badshah and Shamshirband \cite{prop90} & Performance based Service Level Agreement was investigated to optimize all the related parameters for revenue maximization. They worked on customers' satisfaction through  dynamic prices and resources scale-ability.   &  The challenge with this  framework is that performance increases the service cost.   \\

Badshah and Ghani \cite{prop91} &  Resources scalability was explored to maximize the provider revenue. They hired external resources to meet the customers' requirements.   &  The challenge with this is that hiring external resources may decrease the performance and security which may lead to customers' dis-satisfaction.    \\

Hong Zhang and Liu \cite{prop92} &  They worked on revenue maximization by online auction for heterogeneous users' demands.   Their approach works on two main functions. The first one is the payment function and the second one is the resource's allocation rule. The payment function works on the request allocation result and submission time. The allocation rule tries to maximize the bidder's utility.   &  The challenge with this is that customers do not trust  online auctions.      \\

\bottomrule
%\vspace{10mm}%
\end{longtable}
%///////////////////////////////////////////////////////////////////////////////////////////////////////////

\section{Challenges and research directions}
Concerning the IaaS providers’ revenue maximization, penalties and customers' dissatisfaction play a critical role. IaaS resources are non storable and wasted if not utilized on time. Maximum revenue can be earned through maximum utilization.  In the second scenario, most of the providers' revenue is wasted in penalties payment. Cloud providers may also lose dissatisfied customers.  Customer dissatisfaction and rejection leads to loss of revenue. There are many opportunities for revenue maximization in the cloud. Cloud services are available everywhere and at every time i.e. it ubiquitous. Therefore, it is the only business which is in every one hand. If it is run with proper care, it may generate more revenue than other businesses.  To maximize the IaaS cloud providers' revenue,  important challenges and research directions for each of the evaluation criteria are discussed next.

\begin{table*}[ht]
\begin{center}

\caption{Comparative analysis of related studies}
\label{tab:analysis} 

\newcolumntype{b}{X}
\newcolumntype{s}{>{\hsize=.2\hsize}X}
\setlength{\extrarowheight}{3pt}%
\begin{tabularx}{\textwidth}{b s s s s s s s}
\toprule
\textbf{Paper and Authors } & \textbf{C1} & \textbf{C2} & \textbf{C3} & \textbf{C4} & \textbf{C5} & \textbf{C6} & \textbf{C7} \\ 

\midrule

Kundu et al. \cite{Prop52} &     {\xmark}   &	{\cmark}	&	{\xmark}	&	{\xmark}	&	{\cmark}	&	{\xmark}	&	{\xmark}	  \\ 

Macas et al.  \cite{prop31} &  {\cmark}  & {\cmark}  & {\xmark} & {\xmark} & {\xmark} & {\xmark}  &  {\xmark} \\

Gao et al. \cite{Prop38} &  {\xmark}  & {\xmark} & {\xmark} & {\xmark}  & {\cmark} & {\cmark} & {\xmark}  \\

Hadji and Zeghlache \cite{Prop43} &  {\xmark} & {\xmark} & {\cmark} & {\xmark} & {\cmark} & {\xmark} & {\xmark}\\

Amit et al.  \cite{prop83} &   {\xmark}  & {\xmark} & {\cmark} & {\xmark}  & {\cmark} & {\xmark} & {\xmark}  \\

Hamsanandhini and Mohana \cite{prop65}  &  {\xmark}  & {\xmark} & {\cmark} & {\xmark}  & {\xmark} & {\cmark} & {\xmark}  \\

Mei et al. \cite{prop63} & {\xmark}  & {\cmark} & {\xmark} & {\xmark} & {\xmark} & {\xmark} & {\xmark}\\ 

Toosi et al. \cite{prop42} & {\xmark}  & {\cmark} & {\xmark} & {\cmark} & {\cmark} & {\cmark} & {\xmark}\\ 

Zhao et al. \cite{prop56} & {\xmark}  & {\cmark}  & {\xmark} & {\xmark} & {\xmark} & {\cmark} & {\xmark} \\ 

Tevi Yombame Lawson \cite{prop80}  &  {\xmark}  & {\xmark}	& {\xmark} & {\xmark} & {\xmark} & {\cmark} &  {\xmark} \\

Mehiar Dabbagh and Rayes \cite{Prop81} &  {\xmark}  & {\xmark} & {\cmark} & {\xmark}  & {\xmark} & {\cmark} & {\xmark}  \\

Rampersaud and State \cite{prop82} &  {\xmark}  & {\xmark} & {\cmark} & {\xmark}  & {\xmark} & {\cmark} & {\xmark}  \\

Snehanshu Saha and Roy \cite{Prop93} &   {\xmark}  & {\xmark} & {\xmark} & {\xmark}  & {\xmark} & {\cmark} & {\xmark} \\

Hong Xu \cite{Prop78}  &  {\xmark}  &  {\xmark} &  {\xmark} &  {\xmark} & {\cmark} & {\cmark} & {\xmark} \\ 

Adil Maarouf and
Haqiq \cite{Prop75} &  {\cmark} &	{\cmark}	&	{\cmark} 	&	{\xmark}	&	{\xmark}	&	{\xmark}	&	{\xmark}	\\ 

Qi Zhang \cite{Prop77} &  {\xmark}  & {\xmark} & {\cmark} & {\xmark}  & {\cmark} & {\xmark} & {\cmark}  \\ 

Wu and Buyya \cite{Prop34}  &   {\cmark}  & {\cmark} & {\cmark} & {\xmark}  & {\cmark} & {\xmark} & {\xmark}  \\ 

Afzal Badshah and Shamshirband \cite{prop90} &  {\cmark}  & {\cmark} & {\cmark} & {\cmark}  & {\cmark} & {\cmark} & {\xmark}  \\

Badshah and Ghani \cite{prop91} &  {\cmark}  & {\cmark} & {\cmark} & {\cmark}  & {\cmark} & {\cmark} & {\xmark}  \\

Hong Zhang and Liu  \cite{prop92} &   {\xmark}  & {\xmark} & {\xmark} & {\xmark}  & {\xmark} & {\cmark} & {\cmark} \\

\bottomrule
\end{tabularx}
\end{center}
\end{table*}

%//////////////////////////////////////////////////////////

\begin{table*}[ht]
\begin{center}

\caption{Year and approach wise summary of literature}
\label{tab:yearwise} 

\newcolumntype{b}{X}
\newcolumntype{s}{>{\hsize=.2\hsize}X}
\setlength{\extrarowheight}{3pt}%
\begin{tabularx}{\textwidth}{b s b  s}

\toprule
\textbf{Paper and Authors} & \textbf{Pub. Year}  &  \textbf{Approach used} & \textbf{Area}   \\ 
\midrule

Qi Zhang \cite{Prop77} & 2012 & Dynamic Resources Allocation & C5 	\\ 

Wu and Buyya \cite{Prop34} & 2012 & SLA management & C2 	\\

Hong Zhang and Liu \cite{prop92} & 2013 & Online Auction & C7 	\\

Rongdong Hu \cite{prop61} & 2014 & Resources Provisioning & C5 	\\

Hamsanandhini and Mohana \cite{prop65} & 2015 & Client Classification & C4 	\\
	
Adil Maarouf and Haqiq \cite{Prop75} &  2015 & Novel penalty model &   C2 	\\ 

Kundu et al. \cite{Prop52} & 2015 & Resource Management Framework & C1	\\ 

Toosi et al. \cite{prop42} & 2015 & admission control for reservation contracts & C6  \\ 

Hadji and Zeghlache \cite{Prop43} & 2015 & Cloud Federation & C5 	\\ 

Snehanshu Saha and Roy \cite{Prop93} & 2015 & Cost management & C6 	\\

Tevi Yombame Lawson \cite{prop80}  & 2016 & Power Consumption Minimization  & C6 \\

Zhao et al. \cite{prop56} & 2016 & Optimization Scheduling Algorithm & C5 	\\

Mei et al. \cite{prop63} & 2017 & Customer Satisfaction & C4 	\\

Gao et al. \cite{Prop38} & 2018 & Transcoding in Clouds & C1 	\\

Mehiar Dabbagh and Rayes \cite{Prop81} & 2018 & Price Heterogeneity & C6	\\

Rampersaud and State \cite{prop82} & 2019 & Greedy Approximation Algorithm & C5	\\
    
Hou Deng \cite{prop84} & 2019 & Resources Scalability & C3	\\

Afzal Badshah et al. \cite{prop90} & 2019 & Performance based SLA & C2 	\\

Snehanshu Saha and Roy \cite{prop83} & 2019 & Services Performance & C3	\\

Afzal Badshah et al. \cite{prop91} & 2019 &  Resources Scalability  & C3	\\ 

Jing, Mei et al. \cite{prop113} & 2019 &  Dynamic pricing  & C6	\\ 

Benay Kumar et al. \cite{prop114} & 2019 &  Cloud Federation  & C3	\\ 

Ing-Ray et al. \cite{prop139} & 2019 &  Customer trust  & C4	\\

Mengdi   et al. \cite{prop156} & 2019 &  Over booking  & C3	\\

George et al.  \cite{prop157} & 2019 &  Cloud Fedration  & C3	\\

\bottomrule
\end{tabularx}
\end{center}
%\end{small}
\end{table*}

\begin{table*}[h]
\begin{center}

\caption{Potential and challenges of  concern parameters}
\label{tab:pl}

\newcolumntype{c}{X}
\newcolumntype{s}{>{\hsize=.2\hsize}X}
\setlength{\extrarowheight}{1.1pt}%
\newcolumntype{b}{>{\hsize=.5\hsize}X}

\begin{tabularx}{\textwidth}{b c b b c}

\toprule
\textbf{Evaluation Criteria} &  \textbf{Research objectives} & \textbf{Related work} & \textbf{Challanges} & \textbf{Potential solution}\\

C1 \newline Performance management & If considered various performance parameters such as waiting time, running time, response time, security, reliability and availability  &  \cite{Prop59} \cite{Prop50} \cite{Prop52} \cite{Prop98} \cite{Prop108} \cite{Prop109} \cite{Prop110} \cite{Prop111} \cite{Prop112}  & In heavy load, issues in  timing, reliability and availability &   Resources scalability and Performance based Service Level Agreement (PSLA) are a good steps toward performance management \cite{prop90}.\\

C2 \newline  SLA and penalties management &  If acquired  a clear cut SLA to provide above agreed performance parameters according to Agreed SLA & \cite{prop31} \cite{prop73} \cite{prop40} \cite{Prop100} \cite{Prop101} \cite{Prop102} \cite{Prop103} \cite{Prop104} \cite{Prop106} \cite{Prop107} & Not filling the agreed SLA parameters in heavy load & Automatic SLA   worked toward Performance based Service level agreement \cite{prop91}.  \\

C3 \newline  Resources scalability & If questioned the provider about its resources to execute higher load SLAs & \cite{Prop38} \cite{Prop74} \cite{Prop43} \cite{Prop48} \cite{Prop44}  \cite{prop83} \cite{prop84} & Limited resources. Cancellation of SLA in heavy load &  Hiring external resources are getting attached with federated cloud \cite{prop90} \\

C4 \newline  Customer satisfaction & If discussed the customers' satisfaction in terms of customer attraction and retention & \cite{prop65} \cite{prop64} \cite{prop63} \cite{Prop95} \cite{Prop96} \cite{Prop98} \cite{Prop99}   & Dissatisfaction in extreme load and prices &  Offering dynamic prices on customer choices. High performance services.  Authors in \cite{prop91} worked toward customers' satisfaction. \\

C5 \newline  Resources utilization and management & If considered the total resources in use with respect to total available resources & \cite{prop51} \cite{prop62} \cite{prop66} \cite{Prop38} \cite{Prop45} \cite{Prop46} \cite{Prop43}  & Lower utilization of resources and wastage &  Customers' satisfaction, customers, attraction and retention may increase the resources utilization \\

C6 \newline  Cost and prices management & If considered cost minimization by using various methods and reliability of prices & \cite{Prop59}\cite{prop49} \cite{prop42} \cite{prop67} \cite{prop68} \cite{prop71} \cite{prop56} \cite{prop80} \cite{prop69} \cite{Prop81} \cite{prop82} & High cost  so high prices. Wastage of costs on physical and human resources.  & Cos may be reduce by various methods. Suitable prices attract more customers. Dynamic pricing is the best solution for prices issues. \\

C7 \newline  advertisement and auctions & If considered  different parameters such as to reach new customers, to get good auction and also sell underutilized resources.  & \cite{prop57} \cite{prop58} \cite{prop60}  \cite{prop53} \cite{prop92} \cite{Prop94} \cite{Prop95}  & Minimum attention toward new customer attraction & To attract new customers and to sell under utilized resources it is better to do advertisement and auction. \\

\bottomrule
\end{tabularx}
\end{center}
\end{table*}

\subsection{Performance management}
Several research studies reviewed performance for revenue maximization. Ran and Xi \cite{Prop59} used QoS constrains, Feng and Buyya \cite{Prop50} used efficient resources allocation.  Kundu et al. \cite{Prop52} worked on  revenue driven resources allocation. The major limitation  here is the performance degradation due to limited resources or higher utilization. They did not discuss any proper framework to maximize the scalability of the resources according to the incoming requests. It is  known that revenue , performance and cost are related as follows.

\begin{center}
\begin{equation}
 Rev  \propto Per
\end{equation}
\end{center}

where

\begin{center}
\begin{equation}
 Per  \propto Cost
\end{equation}
\end{center}

and 

\begin{center}
\begin{equation}
 Cost  \propto 1/Rev
\end{equation}
\end{center}

The  shows that  $Rev$  increases with  the increase in performance ($Per$).  However, performance needs more cost while cost decreases the revenue.  Since performance, revenues and costs are closely related, there must be a framework for handling this case.  There must be a framework for handling this case. In addition, cloud performance directly depends on the scalability of resources. With higher usage, a server does not work according to the workload requirements. Source scalability and clear SLA can protect the supplier against performance degradation. 

\subsection{SLA and penalties management}
Macas et al. \cite{prop31} explored SLA violation and  cancellation. Emeakaroha et al. \cite{prop40} investigated the lower SLA metrics (e.g., to convert it to higher metrics) to be measured. The main challenges  here are the revenue wastage in penalties payments and the SLA rejection  in extreme utilization. Penalties greatly affect the cloud business. Usually cloud computing accepts loaded SLAs but later on, they cannot provide resources as per the agreement and as a result, they have to pay much of their revenue in penalties.

\begin{center}
\begin{equation}
\eta  \propto V_{n}
\end{equation}
\end{center}

This affects profits as follows.

\begin{center}
\begin{equation}
Prof  = Rev - \eta
\end{equation}
\end{center}

The above expression shows that penalties ($\eta$) increases as the SLA violations ( $V_{n}$) increases. As these penalties are subtracted   from  ($Rev$), therefore, it badly  affect the provider profit ($Prof$).
For such a scenario,  it is extremely necessary to have a clear-cut SLA to provide resources according to agreed parameters. Performance-based Service Level Agreement \cite{prop90},  is a good step toward SLA and penalties management.  

\subsection{Resources scalability}
Gao et al. \cite{Prop38} proposed  \textbf{a} cloud bank to make the resources scalable according to the incoming requests,  Jennifer Ortiz and Balazinska \cite{Prop74} worked on a Personalized Services Level Agreement (PSLA) to provide the services according to a customers' demands from  one place, Hadji and Zeghlache \cite{Prop43} used insourcing and outsourcing techniques in a federated cloud to make the resources scalable. Lakkadwala \cite{Prop48} utilized the migration techniques, Li et al. \cite{Prop44} advances the concept to run private cloud resources on public cloud, Mansour et al. \cite{Prop45} further explored the live cloud migration, Santikarama and Arman \cite{Prop46} used the Economically Customers Relationship Management techniques, Hadji and Zeghlache \cite{Prop43}  worked on live cloud migration, Samanta and Chang \cite{prop83}  worked on resources scalability for mobile applications, and Hou Deng \cite{prop84}  worked to scale the resources using geographically distance servers. 
It is known that Customer Satisfaction (CS) is proportional to resources scalability (SS):

 \begin{center}
\begin{equation}
CS\propto SS 
\end{equation}
\end{center}

This means that  customer satisfaction  increases the provider $Rev$ by the following relation:

 \begin{center}
\begin{equation}
Rev\propto SCS 
\end{equation}
\end{center}

The above expression shows that resources scalability ($SS$) increase the  customer satisfaction which increases the provider ($Rev$).

Services non scalability is the main barrier to increase the cloud business. Providers accept SLAs according to their underlying resources and reject the higher workload. Scalability issues may be partially covered with federated cloud but in federated, insourcing and outsourcing only can be done with registered groups. This issue may be covered by hiring external resources.  In such a case two SLAs are implemented. The first SLA is negotiated between  Cloud Service Provider (CSP) and customer, while the second  SLA is negotiated between  Cloud Service Provider (CSP) and External Cloud Service Provider (ECSP). External resources also affect performance and security.  Cloud providers should not hire complete resources from external cloud service because they pay to external cloud services according to the usage of services.

\subsection{Customer satisfaction}
Hamsanandhini and Mohana \cite{prop65} investigated customers' satisfaction for revenue maximization, Manzoor et al. \cite{prop64} used the customers centered approach, and Mei et al. \cite{prop63} discussed the customers' satisfaction by full filing the Quality of Services (QoS) and Prices of Services (PoS) parameters.
The following expression shows that customer satisfaction ($CS$) depends on services scalability ($SS$), services efficiency ($Eff$), and Quality of Services ($QoS$).

 \begin{center}
\begin{equation}
CS\propto SS \times Eff \times QoS
\end{equation}
\end{center}

Customer satisfaction can be increased by providing good QoS. Good QoS requires scalable resources. Prices also have a major impact on customers. Some customers prefer performance, while others prefer lower prices. A good price framework can affect more customers. Customer support is the most important reason for customer satisfaction because they feel confident with the right customer support.

\subsection{Resources utilization and management}
 Shin et al. \cite{prop51} further investigated the deadline guaranteed resources utilization, Balagoni and Rao \cite{prop62} discussed the scheduling policies for heterogeneous clouds, Yuan et al. \cite{prop66} proposed temporal task scheduling in the hybrid cloud, and Gao et al. \cite{Prop38} worked on transcoding video streaming. With limited resources and maximum resource utilization, the major problem  here is that cloud providers reject  existing customers  if their  penalties are lower than new customers' revenue. There are different QoS SLAs, the combination of those SLAs is adopted which bring a higher revenue and lower revenue SLAs are cancelled. 
 
Rejecting any customer is a great loss in the cloud business. Such providers will never be trusted in the future. Resources utilization discusses the total revenue earned by total available resources.  Efficient resource utilization depends on customer satisfaction, attraction, retention and accepting dynamic SLAs. Working on the resources scalability and dynamic prices may increase resource utilization. 

\subsection{Cost and prices management}
Ran and Xi \cite{Prop59}  explored the dynamic pricing model in cloud computing, Zhang and Boutaba \cite{prop49} used market analyzer, capacity planer and dynamic pricing scheme,  Toosi et al. \cite{prop42} used optimal capacity and different pricing schemes, Chi et al.  \cite{prop67} used efficient resources scheduling and prices models, Zhou et al.  \cite{prop68} worked on cost optimization , Ibrahim et al  \cite{prop71} worked on hybrid cost and priority-based scheduling,  Tevi Yombame Lawson \cite{prop80} worked on cost minimization to maximize the profit,  Tang and Chen \cite{prop69} proposed economic framework for resources management, Mehiar Dabbagh and Rayes  \cite{Prop81} worked on prices and capacity planning and Rampersaud and State \cite{prop82} worked on efficient resources allocation and costing. 

Prices are fixed according to the total cost. Cost may be minimized by efficiently managing power consumption and human resources.  Joint prices may be used for cloud business. Fixed rates may be used for high-performance customers. Spot pricing may be used for under-utilized resources. It will increase resource utilization.  The total revenue is :

\begin{center}	
 \begin{equation}
Rev_{total} =  Rev_{res} + Rev_{spot} + Rev_{unit} 
  \end{equation}
		\end{center}

where $Rev_{res}$  is the revenue from reserved resources, $Rev_{spot}$ are earned from Spot pricing and  $Rev_{unit}$ is the revenue earned from unit based charges.  The total profit is:

\begin{center}	
 \begin{equation}
prof_{total} =   \sum_{k=0}^{n} Rev -  \chi_{total}
  \end{equation}
		\end{center}

Where  $prof_{total}$ is the total profit earned and $\chi_{total}$ is total cost. 

\subsection{Advertisement and over-utilization}
Dabbagh et al. \cite{prop57} integrated and further investigated the over-commitment techniques to keep the resources busy, Metwally and Ahmed \cite{prop58} used the resources optimization tool, Hammoudi
et al. \cite{prop60} used the multi-agent architecture for load balancing,  and Deng et al. \cite{prop53} worked on the online auction. 

The major challenge  here is that advertisements may require a high cost. This will have a major impact on cloud profit. Recent marketing and advertisement techniques may be used to reach and attract new customers.

\section{Conclusion} 
This article presents the concepts of IaaS clouds, the challenges, and opportunities towards revenue maximization. It further explores the different techniques used to maximize the providers' revenue. This study provides a good foundation to scholars and practitioners who are interested to work in IaaS clouds revenue maximization.  During the literature study, we came across several challenges.  Performance drops due to several reasons. Resources non-scalability is one of them. Performance degradation violates SLA. It is the major challenge towards IaaS providers' revenue. In the case of a high workload, providers pay the revenue back in penalties instead of making  profit. Cloud Service Providers (CSP) are virtual in the cloud business. Customers hesitate to trust the virtual provider for performance, privacy, and security. Due to the massive workloads, it is also challenging to properly utilize, schedule and migrate the workload on the resources. 

We reviewed the revenue maximization techniques in detail and classified it into different categories. Performance management covers the running time, response time, bandwidth, resources availability, and reliability. SLA and penalties management discussed the agreed terms for the above performance parameters. Resources scalability addresses the capacity of the provider to entertain heavy dynamic customers' demands.  Customer satisfaction covers the techniques for customer retention.  Resources utilization and management discussed all the scheduling and migration policies to efficiently schedule and utilize the underlying resources. Cost and prices management covered how to determine the prices of the resources by analyzing the total cost and margin. By advertisement, the provider may get a handsome workload to keep its resources utilized.

Using these parameters as a base (as shown in Table \ref{tab:criteria}), one criterion was introduced and all studies were critically reviewed accordingly. Subsequently, we presented the strengths of revenue maximization in the cloud business. Performance of the cloud resources may be increased by using Performance-based Service Level Agreement. To cover the resources non-scalability, external resources may be hired. Dynamic prices attract more customers which  increases customers' satisfaction and resources utilization. Finally,  we tabulated all the opportunities of every category of the cloud business to make it easy for readers.

%\nocite{*}
%\bibliographystyle{WileyNJD-AMA}
\bibliography{wileyNJD-AMA}

\begin{thebibliography}{100}
\providecommand \doibase [0]{http://dx.doi.org/}%

\bibitem{Prop149}
Davis G. 2020: Life with 50 billion connected devices. In:  {\it IEEE
  International Conference on Consumer Electronics (ICCE)}IEEE. ; 2018; Las
  Vegas, NV, USA\string: 1-1

\bibitem{Prop17}
Forbes: Cloud computing forecast.
  \url{https://www.forbes.com/sites/louiscolumbus/2017/04/29/roundup-of-cloud-computing-forecasts2017/#5c42322c31e8/};
  2017.

\bibitem{Prop154}
Annual forecast of data-sphere.
  \url{www.forbes.com/sites/tomcoughlin/2018/11/27/175-zettabytes-by-2025/#59c815325459//};
  2019.

\bibitem{Prop04}
Wang X, Wang B, Huang J. Cloud computing and its key techniques Computer
  Science and Automation Engineering (CSAE). In:  {\it International Conference
  on Cloud Computing}IEEE. ; 2011; Shanghai, China\string: 404-410.

\bibitem{Prop07}
Lei X, Liao X, Huang T, Li H. Cloud computing service: The caseof large matrix
  determinant computation. {\it IEEE Transactions on Services Computing}
  2015\string; 8(5)\string: 688-700.

\bibitem{Prop08}
Ziglari H, Yahya S. Deployment models: Enhancing security in cloud computing
  environment. In:  {\it 22nd Asia-Pacific Conference on Communications
  (APCC)}IEEE. IEEE; 2016; Shanghai, China\string: 1-6.

\bibitem{Prop19}
Amazon Web Services. \url{http://aws.amazon.com/ec2/pricing/};  2019.

\bibitem{Prop123}
Microsoft Azure Cloud. \url{http://azure.microsoft.com/en-us/};  2019.

\bibitem{Prop124}
Google Cloud Computing Services. \url{https://cloud.google.com/};  2019.

\bibitem{Prop125}
Alibaba Cloud. \url{https://www.alibabacloud.com/};  2019.

\bibitem{Prop126}
IBM Cloud. \url{https://www.ibm.com/cloud/};  2019.

\bibitem{Prop127}
Oracle Cloud. \url{https://cloud.oracle.com/home};  2019.

\bibitem{Prop128}
SAP Cloud Platform. \url{https://cloudplatform.sap.com/index.html/};  2019.

\bibitem{Prop129}
Cobweb Cloud Solutions. \url{https://cobweb.com/};  2019.

\bibitem{Prop130}
Mulesoft Cloud Computing Services.
  \url{https://www.mulesoft.com/resources/cloudhub/cloud-computing-service};
  2019.

\bibitem{Prop131}
Salesforce Cloud Computing Services.
  \url{https://www.salesforce.com/eu/products/what-is-salesforce/};  2019.

\bibitem{Prop02}
Buyya R, Ranjan R, Calheiros R. Intercloud: Utility-oriented federation of
  cloud computing environments for scaling of application services. {\it
  Algorithms and architectures for parallel processing} 2010\string; 4\string:
  13--31.

\bibitem{Prop03}
Ateeqa J, Afzal B, Tauseef R. SLA based Infrastructure resources allocation in
  Cloud computing to increase IaaS provider revenue. {\it Research Journal of
  Science and IT Management} 2015\string; 4(3)\string: 37-44.

\bibitem{Prop11}
Zhou AC, He B, Liu C. Monetary cost optimizations for hosting
  workflow-as-a-service in IaaS clouds. {\it IEEE transactions on cloud
  computing} 2016\string; 4(1)\string: 34--48.

\bibitem{Prop13}
Matrazali N, H. M, Uda R, Kinoshita T, Shiratori M. Vulnerability analysis
  using network timestamps in full virtualization virtual machine. In:  {\it
  International Conference on Information Systems Security and Privacy
  (ICISSP)}IEEE. ; 2015; Angers, France\string: 83-89.

\bibitem{Prop14}
Kaushik A, Chaturvedi A. A review of efficient data utilization schemes in
  cloud computing. In:  {\it 3rd International Conference on Computing for
  Sustainable Global Development (INDIACom)}IEEE. ; 2016; New Delhi,
  India\string: 1--1.

\bibitem{Prop145}
McKinsey Theory. \url{https://beyondphilosophy.com//};  2019.

\bibitem{Prop99}
Raqual.~A. A, Glaucu.~H. M, Gilberto. MDG. Factors Influencing Customer
  Satisfaction in Software as a Service (SaaS): Proposal of a System of
  Performance Indicators. {\it IEEE LATIN AMERICA TRANSACTIONS} 2017\string;
  15(8)\string: 1536-1541.

\bibitem{Prop81}
Mehiar D, Bechir H, Mohsen G, Ammar R. Exploiting Task Elasticity and Price
  Heterogeneity for Maximizing Cloud Computing Profits. {\it IEEE Transactions
  on Emerging Topics in Computing} 2018\string; 6(1)\string: 85 - 96.

\bibitem{Prop116}
Qiang D. Cloud service performance evaluation: status, challenges, and
  opportunities -- a survey from the system modeling perspective. {\it Digital
  Communications and Networks} 2017\string; 3(2)\string: 101-111.

\bibitem{Prop118}
Hero M, Rosli S, Amir H. A Survey on Cloud Computing Security. {\it Archives
  des Science} 2012\string; 56(6)\string: 2-9.

\bibitem{Prop115}
Anup~H. G. A Survey paper on Cloud Computing and its effective utilization with
  Virtualization. {\it International Journal of Scientific and Engineering
  Research} 2013\string; 4(12)\string: 363--375.

\bibitem{Prop119}
PritiKumari P. A survey of fault tolerance in cloud computing. {\it Journal of
  King Saud University - Computer and Information Sciences} 2018\string: 1-18.
\newblock \href {\doibase https://doi.org/10.1016/j.jksuci.2018.09.021} {doi:
  https://doi.org/10.1016/j.jksuci.2018.09.021}

\bibitem{Prop120}
Shah JM, Kotecha K, Pandya S, Choksi D, Joshi N. Load Balancing in cloud
  computing: Methodological Survey on different types of algorithm. In:  {\it
  International Conference on Trends in Electronics and Informatics
  (ICEI)}IEEE. ; 2017\string: 100-107.

\bibitem{Prop121}
Ma W, Zhang J. The survey and research on application of cloud computing. In:
  {\it 7th International Conference on Computer Science \& Education
  (ICCSE)}IEEE. ; 2012\string: 203--206.

\bibitem{Prop159}
Mehdi S, Hamid T, Ejaz A, Abdullah G, Muhammad~Khurram K. A review on remote
  data auditing in single cloud server: Taxonomy and open issues. {\it Journal
  of Networks and Computer Applications} 2014\string; 43(14)\string: 121-141.

\bibitem{Prop06}
The Cloud Service Industrys 10 Most Critical Metrics.
  https://guidingmetrics.com/content/cloud-services-industrys-10-most-critical-metrics/;
  2019.

\bibitem{Prop09}
Buyya R, Broberg J, Goscinski AM. {\it Cloud computing: Principles and
  paradigms}. 87.
\newblock John Wiley \& Sons .
\newblock 2010.

\bibitem{Prop152}
Lei Z, Anmin F, Shui Y, Mang S, , Boyu K. Data integrity verification of the
  outsourced big data in the cloud environment: A survey. {\it Journal of
  Network and Computer Applications} 2019\string; 112\string: 1-15.

\bibitem{Prop158}
Sayed~Chhattan S, Sajad~Hussain C, Ali~Kashif B. A Centralized Location-Based
  Job Scheduling Algorithm for Inter-Dependent Jobs in Mobile Ad Hoc
  Computational Grids. {\it Journal of Applied Sciences} 2010\string;
  10(3)\string: 174-181.

\bibitem{Prop21}
Nguyen NC, Wang P, Niyato D, Wen Y, Han Z. Resource management in cloud
  networking using economic analysis and pricing models: a survey. {\it IEEE
  Communications Surveys \& Tutorials} 2017\string; 9(2)\string: 954-1001.

\bibitem{Prop140}
Siqian G, Beibei Y, Zheng Z, Kai-Yuan C. Adaptive Multivariable Control for
  Multiple Resource Allocation of Service-Based Systems in Cloud Computing.
  {\it IEEE Access} 2019\string; 16\string: 13817-13831.

\bibitem{Prop150}
Masoud A, Saleh Y, Dusit N. Pricing strategies of IoT wide area network service
  providers with complementary services included. {\it Journal of Network and
  Computer Applications} 2019\string; 147\string: 102426.
\newblock \href {\doibase https://doi.org/10.1016/j.jnca.2019.102426} {doi:
  https://doi.org/10.1016/j.jnca.2019.102426}

\bibitem{Prop151}
Ankita A, Gregory V, Seghbroecka H, Morab F, De T, Bruno V. SpeCH: A scalable
  framework for data placement of data-intensive services in geo-distributed
  clouds. {\it Journal of Network and Computer Applications} 2019\string;
  142\string: 1-14.

\bibitem{Prop22}
Sinung S, Suhono S, Suhardi , Roberdm S. PERFORMANCE MEASUREMENT OF CLOUD
  COMPUTING SERVICES. {\it International Journal on Cloud Computing: Services
  and Architecture(IJCCSA)} 2012\string; 2(2)\string: 9-22.

\bibitem{Prop59}
Ran Y, Jian Z, Xi H. Dynamic iaas computing resource provisioning strategy with
  qos constraint. {\it IEEE Transactions on Services Computing} 2015\string;
  10(2)\string: 190-202.

\bibitem{Prop97}
Danilo A, Giuliano C, Michele C. Quality-of-service in cloud computing:
  modeling techniques and their applications. {\it Journal of Internet Services
  and Applications} 2013\string; 5(1)\string: 5-11.

\bibitem{Prop52}
Kundu S, Rangaswami R, Zhao M, Gulati A, Dutta K. Revenue Driven Resource
  Allocation for Virtualized Data Centers. In:  {\it International Conference
  on Autonomic Computing (ICAC)}IEEE. ; 2015\string: 12-21.

\bibitem{Prop50}
Feng G, Buyya R. Maximum revenue-oriented resource allocation in cloud. {\it
  International Journal of Grid and Utility Computing} 2016\string;
  7(1)\string: 12-21.

\bibitem{Prop98}
Nazanin P, Abbas T, Sanaei M. A model for evaluating cloud-computing users
  satisfaction. {\it African Journal of Business Management} 2013\string;
  7(16)\string: 1405-1413.

\bibitem{Prop108}
Ioannis G, Dimitrios T, Nectarios K. Towards an Adaptive, Fully Automated
  Performance Modeling Methodology for Cloud Applications. In:  {\it IEEE
  International Conference on Cloud Engineering (IC2E)}IEEE. ; 2018; Orlando,
  FL, USA\string: 148-158

\bibitem{Prop109}
Dung N, Andre L, Edward D, Ken K, Amy A. Evaluation of Highly Available Cloud
  Streaming Systems for Performance and Price. In:  {\it 18th IEEE/ACM
  International Symposium on Cluster, Cloud and Grid Computing (CCGRID)}IEEE. ;
  2018; Washington, DC, USA\string: 360-363

\bibitem{Prop110}
Bauer E. Cloud Automation and Economic Efficiency. {\it IEEE Cloud Computing}
  2018\string; 5(2)\string: 26-32.

\bibitem{Prop111}
Vladimir P, Anshul J, Michael G. IaaS Reactive Autoscaling Performance
  Challenges. In:  {\it IEEE 11th International Conference on Cloud Computing
  (CLOUD)}IEEE. ; 2018; San Francisco, CA, USA\string: 954-957

\bibitem{Prop112}
Songtai D, Ao Z, Shangguang W. The Performance Evaluation of Virtual Machine
  Placement Algorithm Based on WebCloudSim. In:  {\it IEEE 11th International
  Conference on Cloud Computing (CLOUD)}. 1. IEEE. ; 2018\string: 950-953.

\bibitem{prop31}
Macas M, Fit JO, Guitart J. Rule-based {SLA} management for revenue
  maximization in cloud computing markets. In:  {\it International Conference
  on Network and Service Management (CNSM)}IEEE. ; 2010; Niagara Falls,
  Canada\string: 354--357

\bibitem{prop35}
Wu L, Garg SK, Buyya R. {SLA}-based resource allocation for software as a
  service provider (saas) in cloud computing environments. In:  {\it 11th
  International Symposium on Cluster, Cloud and Grid Computing (CCGrid)}IEEE.
  IEEE; 2011; Newport Beach, CA, USA\string: 195--204

\bibitem{prop73}
Christpher R, Ivan~Breskovic I, Schahram . Automatic {SLA} Matching and
  Provider selection in Grid and Cloud Computing Environments. In:  {\it GRID
  '12 Proceedings of the ACM/IEEE 13th International Conference on Grid
  Computing}ACM. ; 2012; Las Vegas, NV, USA\string: 85--94

\bibitem{Prop74}
Jennifer O, Victor TA, Balazinska M. A vision for personalized service level
  agreements in the cloud. In:  {\it Proceedings of the Second Workshop on Data
  Analytics in the Cloud}ACM. ; 2013\string: 1-5.

\bibitem{prop40}
Emeakaroha V, Brandic I, Maurer M, Dustdar S. Low level metrics to high level
  SLAs-LoM2HiS framework: Bridging the gap between monitored metrics and SLA
  parameters in cloud environments. In:  {\it International Conference on High
  Performance Computing and Simulation (HPCS)}IEEE. ; 2010; Caen,
  France\string: 48-54

\bibitem{Prop105}
Catello DM, Santonu S, Rajeshwari G, Zbigniew~T. K, Ravishankar~K. I. Analysis
  and Diagnosis of SLA Violations in a Production SaaS Cloud. {\it IEEE
  Transactions on Reliability} 2017\string; 66(1)\string: 54-75.

\bibitem{Prop100}
Shivani , Ajmer S. Taxonomy of SLA violation minimization techniques in cloud
  computing. In:  {\it Second International Conference on Inventive
  Communication and Computational Technologies (ICICCT)}IEEE. ; 2018;
  Coimbatore, India\string: 1845-1850

\bibitem{Prop101}
Shahin V, Catherine~Truchan JK, Halima E. Automated Enforcement of SLA for
  Cloud Services. In:  {\it IEEE 11th International Conference on Cloud
  Computing (CLOUD)}IEEE. ; 2018; San Francisco, CA, USA\string: 49-56

\bibitem{Prop102}
Taher L, Achraf M, Walid G, Samir T, Faiez G. Cloud SLA Modeling and
  Monitoring. In:  {\it IEEE International Conference on Services Computing
  (SCC)}IEEE. ; 2017; Honolulu, HI, USA\string: 338-345

\bibitem{Prop103}
Alayat H, Farookh KH, Omar H, Ravindra B, Elizabeth C, Alexan . Risk-based
  framework for {SLA} violation abatement from the cloud service provider's
  perspective. {\it The Computer Journal} 2017\string; 6(9)\string: 1306-1322.

\bibitem{Prop104}
Rahul Y, Weizhe Z, Omprakash K, Prabhat~Ranjan S, Ibrahim~A. E. Adaptive
  Energy-Aware Algorithms for Minimizing Energy Consumption and SLA Violation
  in Cloud Computing. {\it IEEE Access} 2018\string; 6(9)\string: 55923-55936.

\bibitem{Prop106}
Shengli Z, Lifa W, Jin C. A privacy-based SLA violation detection model for the
  security of cloud computing. {\it China Communications} 2017\string;
  15(9)\string: 155-165.

\bibitem{Prop107}
Yuan X, Li Y, Jia T, Liu T, Wu Z. An Analysis on Availability Commitment and
  Penalty in Cloud SLA. In:  {\it IEEE 39th Annual Computer Software and
  Applications Conference}IEEE. ; 2015; Taichung, Taiwan\string: 914-919

\bibitem{Prop38}
Gao G, Hu H, Wen Y, Westphal C. Resource Provisioning and Profit Maximization
  for Transcoding in Clouds: A Two-Timescale Approach. {\it IEEE Transactions
  on Multimedia} 2017\string; 19(4)\string: 836--848.

\bibitem{Prop43}
Hadji M, Zeghlache D. Mathematical programming approach for revenue
  maximization in cloud federations. {\it IEEE Transactions on Cloud Computing}
  2015\string; 5(1)\string: 99 -111.

\bibitem{Prop48}
Upadhyay A, Lakkadwala P. Migration of over loaded process and schedule for
  resource utilization in Cloud Computing. In:  {\it 4th International
  Conference on Reliability, Infocom Technologies and Optimization
  (ICRITO)(Trends and Future Directions)}IEEE. ; 2015; Noida, India\string: 1-4

\bibitem{Prop44}
Li Y, Zhang J, Hu Q, Pei J. Research and practice on the theory of private
  clouds migration. In:  {\it 13th International Conference on Signal
  Processing (ICSP)}IEEE. ; 2016; Chengdu, China\string: 1813-1818

\bibitem{prop83}
Amit S, Zheng C. Adaptive Service Offloading for Revenue Maximization in Mobile
  Edge Computing with Delay-Constraint. {\it IEEE Internet of Things Journal}
  2019\string; 6(2)\string: 3864 - 3872.

\bibitem{prop84}
Hou D, Liusheng H, Hongli X. Revenue Maximization for Dynamic Expansion of
  Geo-distributed Cloud Data Centers. {\it IEEE Transactions on Cloud
  Computing} 2018\string: 1-13.
\newblock \href {\doibase 10.1109/TCC.2018.2808351} {doi:
  10.1109/TCC.2018.2808351}

\bibitem{prop65}
Hamsanandhini S, Mohana R. Maximizing the revenue with client classification in
  Cloud Computing market. In:  {\it International Conference on Computer
  Communication and Informatics (ICCCI)}IEEE. ; 2015; Coimbatore, India\string:
  1-7

\bibitem{Prop95}
Tram TH, Chen-Khong T. An Auction-based Resource Allocation Model for Green
  Cloud Computing. In:  {\it IEEE International Conference on Cloud
  Engineering}IEEE. ; 2013; Redwood City, CA, USA\string: 269-279

\bibitem{prop64}
Manzoor S, Taha A, Suri N. Trust Validation of Cloud IaaS: A Customer-centric
  Approach. In:  {\it Trustcom/BigDataSE SPA, IEEE}IEEE. ; 2016; Tianjin,
  China\string: 97-104

\bibitem{prop63}
Mei J, Li K, Li K. Customer-Satisfaction-Aware Optimal Multiserver
  Configuration for Profit Maximization in Cloud Computing. {\it IEEE
  Transactions on Sustainable Computing} 2017\string; 2(1)\string: 17--29.

\bibitem{prop51}
Shin S, Kim Y, Lee S. Deadline-guaranteed scheduling algorithm with improved
  resource utilization for cloud computing. In:  {\it Consumer Communications
  and Networking Conference (CCNC)}IEEE. ; 2015; Las Vegas, NV, USA\string:
  814-819

\bibitem{prop62}
Balagoni Y, Rao RR. Locality-Load-Prediction Aware Multi-Objective Task
  Scheduling in the Heterogeneous Cloud Environment. {\it Indian Journal of
  Science and Technology} 2017\string; 10(9)\string: 1-9.

\bibitem{prop66}
Yuan H, Bi J, Tan W, Li BH. Temporal task scheduling with constrained service
  delay for profit maximization in hybrid clouds. {\it IEEE Transactions on
  Automation Science and Engineering} 2017\string; 14(1)\string: 337-348.

\bibitem{prop71}
Ibrahim E, El-Bahnasawy NA, Omara FA. Task Scheduling Algorithm in Cloud
  Computing Environment Based on Cloud Pricing Models. In:  {\it World
  Symposium on Computer Applications \& Research (WSCAR)}IEEE. ; 2016\string:
  65--71.

\bibitem{Prop45}
Mansour I, A. E, Cooper K, Bouchachia H. Effective Live Cloud Migration. In:
  {\it 4th International Conference on Future Internet of Things and Cloud
  (FiCloud)}IEEE. ; 2016; Vienna, Austria\string: 1-1

\bibitem{Prop46}
Santikarama I, Arman AA. Designing enterprise architecture framework for
  non-cloud to cloud migration using TOGAF, CCRM, and CRMM. In:  {\it
  International Conference on ICT For Smart Society (ICISS)}IEEE. ; 2016;
  Surabaya, Indonesia\string: 32-37

\bibitem{Prop47}
Tsakalozos K, Verroios V, Roussopoulos M, Delis A. Live VM Migration under
  Time-Constrains in Share-Nothing IaaS-Clouds. {\it IEEE Transactions on
  Parallel and Distributed Systems} 2017\string; 28(8)\string: 2285-2298.

\bibitem{prop49}
Zhang Q, Boutaba R. Dynamic workload management in heterogeneous cloud
  computing environments. In:  {\it Network Operations and Management Symposium
  (NOMS), IEEE}IEEE. ; 2014; Krakow, Poland\string: 1-1

\bibitem{prop42}
Toosi AN, Vanmechelen K, Ramamohanarao K, Buyya R. Revenue maximization with
  optimal capacity control in infrastructure as a service cloud markets. {\it
  IEEE transactions on Cloud Computing} 2015\string; 3(3)\string: 261--274.

\bibitem{prop67}
Chi Y, Li X, Wang X, Leung VC, Shami A. A Fairness-Aware Pricing Methodology
  for Revenue Enhancement in Service Cloud Infrastructure. {\it IEEE Systems
  Journal} 2015\string; 11(2)\string: 1006 - 1017.

\bibitem{Prop70}
Hong X, Baochun L. Dynamic Cloud Pricing for Revenue Maximization. {\it IEEE
  Transaction on Cloud Computing} 2013\string; 1(2)\string: 158-172.

\bibitem{prop56}
Zhao Y, Calheiros R, Bailey J, Sinnott R. SLA-based profit optimization for
  resource management of big data analytics-as-a-service platforms in cloud
  computing environments. In:  {\it International Conference on Big Data (Big
  Data)}IEEE. ; 2016; Washington, DC, USA\string: 432-441

\bibitem{prop80}
Tevi~Yombame L, Zbigniew D. Economic framework for resource management in data
  centers. In:  {\it IEEE International Conference on Communication Systems
  (ICCS)}ACM. ; 2016; Shenzhen, China\string: 1-4

\bibitem{prop69}
Tang L, Chen H. Joint pricing and capacity planning in the IaaS cloud market.
  {\it IEEE Transactions on Cloud Computing} 2014\string; 5(1)\string: 57 - 70.

\bibitem{prop82}
Safraz R, Wayne S. An Approximation Algorithm for Sharing-Aware Virtual Machine
  Revenue Maximization. In:  {\it IEEE Transactions on Services Computing}IEEE.
  ; 2018\string: 1-1

\bibitem{Prop93}
Snehanshu S, Jyotirmoy S, Avantika D, Ranjan R. A novel revenue optimization
  model to address the operation and maintenance cost of a data center. {\it A
  journal of cloud computing} 2016\string: 1-23.
\newblock \href {\doibase https://doi.org/10.1186/s13677-015-0050-8} {doi:
  https://doi.org/10.1186/s13677-015-0050-8}

\bibitem{prop57}
Dabbagh M, Hamdaoui B, Guizani M, Rayes A. Efficient datacenter resource
  utilization through cloud resource overcommitment. In:  {\it Conference on
  Computer Communications Workshops (INFOCOM WKSHPS)}IEEE. ; 2015; Hong Kong,
  China\string: 330-335

\bibitem{prop58}
Metwally K, Abdallah K, Ahmed . A cost-efficient QoS-aware model for cloud IaaS
  resource allocation in large datacenters. In:  {\it 4th International
  Conference on Cloud Networking (CloudNet)}IEEE. ; 2015; Niagara Falls, ON,
  Canada\string: 38-43

\bibitem{Prop148}
Samimi P, Teimouri Y, Mukhtar M. A combinatorial double auction resource
  allocation model in cloud computing. {\it Information Sciences} 2016\string;
  357\string: 201-216.

\bibitem{prop60}
Hammoudi S, Benaouda A, Harous S, Aliouat Z. Load balancing in the cloud using
  specialization. In:  {\it Ubiquitous Computing, Electronics \& Mobile
  Communication Conference (UEMCON), IEEE Annual}IEEE. ; 2016; New York, NY,
  USA\string: 1-7

\bibitem{Prop78}
Hong X, Baochun L. Maximizing revenue with dynamic cloud pricing: The infinite
  horizon case. In:  {\it IEEE International Conference on Communications
  (ICC)}ACM. ; 2012; Ottawa, ON, Canada\string: 2929-2933

\bibitem{Prop75}
Adil M, Bouchra Eq, Abderrahim M, Abdelkrim H. A Novel Penalty Model for
  Managing and Applying Penalties in Cloud Computing. In:  {\it IEEE ACS 12th
  International Conference of Computer Systems and Applications (AICCSA)}ACM. ;
  2015; Marrakech, Morocco\string: 85-94

\bibitem{Prop77}
Qi Z, Quanyan Z, Raouf B. Dynamic Resource Allocation for Spot Markets in Cloud
  Computing Environments. In:  {\it Fourth IEEE International Conference on
  Utility and Cloud Computing}ACM. ; 2011; Marrakech, Morocco\string: 85-94

\bibitem{Prop34}
Wu L, Buyya R. Service level agreement (SLA) in utility computing systems. {\it
  IGI Global} 2012\string; 15\string: 1-26.

\bibitem{prop90}
Afzal B, Anwar G, Shahaboddin S, Giuseppe A, Antonio P. Performance based
  Service Level Agreement (PSLA) in Cloud Computing to Optimize Penalties and
  Revenue. {\it IET Communications} 2020.
\newblock \href {\doibase 10.1049/iet-com.2019.0855} {doi:
  10.1049/iet-com.2019.0855}

\bibitem{prop91}
Afzal B, Shahaboddin S, Anwar G, Anthony C. Optimizing IaaS Provider Revenue
  through Customer Satisfaction and Efficient Resource Provisioning in Cloud
  Computing. {\it IET Communications} 2019\string; 13(9)\string: 2913--2922.
\newblock \href {\doibase 10.1049/iet-com.2019.0554} {doi:
  10.1049/iet-com.2019.0554}

\bibitem{prop92}
Hong Z, Hongbo J, Bo L. A Framework for Truthful Online Auctions in Cloud
  Computing with Heterogeneous User Demands. {\it IEEE transaction on cloud
  computing} 2016\string; 65(3)\string: 805-818.

\bibitem{prop61}
Rongdong H, Jingfei J, Guangming L. Efficient Resources Provisioning Based on
  Load Forecasting in Cloud. {\it The scientific world journal} 2014\string;
  2014\string: 1-12.
\newblock \href {\doibase http://dx.doi.org/10.1155/2014/321231} {doi:
  http://dx.doi.org/10.1155/2014/321231}

\bibitem{prop113}
Jing M, Kenli L, Zhao T, Qiang L, Keqin L. Profit Maximization for Cloud
  Brokers in Cloud Computing. {\it IEEE Transactions on Parallel and
  Distributed Systems} 2019\string; 30(1)\string: 190 - 203.

\bibitem{prop114}
Benay~Kumar R, Avirup S, Sunirmal K, Sarbani R. Toward maximization of profit
  and quality of cloud federation: solution to cloud federation formation
  problem. {\it The Journal of Super computing} 2019\string; 75(2)\string:
  885--929.

\bibitem{prop139}
Ing-Ray C, Jia G, Ding-Chau W, Jeffrey JPT, Hamid AH, Ilsun Y. Trust-Based
  Service Management for Mobile Cloud IoT Systems. {\it IEEE Transactions on
  Network and Service Management} 2019\string; 16(1)\string: 4294 - 4308.

\bibitem{prop156}
Mengdi Y, Donglin C, Shang J. Optimal Overbooking Policy for Cloud Service
  Providers: Profit and Service Quality. {\it IEEE Access} 2019\string;
  7\string: 96132 - 96147.

\bibitem{prop157}
George D, Iordanis K, George~D S. Cloud Federations: Economics, Games and
  Benefits. {\it IEEE/ACM Transactions on Networking} 2019\string; 27\string:
  2111 - 2124.

\bibitem{Prop96}
Mario M, Jordi G. Client Classification Policies for {SLA} Enforcement in
  Shared Cloud Datacenters. In:  {\it 12th IEEE/ACM International Symposium on
  Cluster, Cloud and Grid Computing}IEEE/ACM. IEEE/ACM; 2012; Ottawa, ON,
  Canada

\bibitem{prop68}
Zhou AC, He B, Liu C. Monetary cost optimizations for hosting
  workflow-as-a-service in IaaS clouds. {\it IEEE transactions on cloud
  computing} 2016\string; 4(1)\string: 34--48.

\bibitem{prop53}
Deng X, Xiao T, Zhu K. Learn to Play Maximum Revenue Auction. In:  {\it IEEE
  Transactions on Cloud Computing}IEEE. IEEE; 2019\string: 1--10.

\bibitem{Prop94}
Wang X, Sun J, Li H, Wu C, Huang M. A reverse auction based allocation
  mechanism in the cloud computing environment. {\it Applied Mathematics \&
  Information Sciences} 2013\string; 7(1)\string: 75-84.

\end{thebibliography}

\end{document}